\documentclass[12pt]{article}
\usepackage{epsfig}
\newcommand{\bmat}{\left(\begin{array}}
\newcommand{\emat}{\end{array}\right)}
\def\NPB#1#2#3{Nucl. Phys. B{#1} (19#2) #3}
\def\PLB#1#2#3{Phys. Lett. B{#1} (19#2) #3}

\def\PRD#1#2#3{Phys. Rev. D{#1} (19#2) #3}

\def\preal{{\rm Re\,}}

\def\yzero{\smash{\hbox{$y\kern-4pt\raise1pt\hbox{${}^\circ$}$}}}

\def\-{\hphantom{-}}

\def\s2{\frac{1}{\sqrt2}}

\def\beq{\begin{equation}}
\def\eeq{\end{equation}}
\def\beqa{\begin{eqnarray}}
\def\eeqa{\end{eqnarray}}

\def\IF{\relax{\rm I\kern-.18em F}}
\def\II{\relax{\rm I\kern-.18em I}}
\def\IP{\relax{\rm I\kern-.18em P}}
\def\IC{\relax\hbox{\kern.25em$\inbar\kern-.3em{\rm C}$}}
\def\IR{\relax{\rm I\kern-.18em R}}

\def\Dsl{\,\raise.15ex\hbox{/}\mkern-13.5mu D} 
\def\IZ{Z\kern-.4em  Z}
\def\bmat{\left(\begin{array}}
\def\emat{\end{array}\right)}

\def    \part          {\partial}
\def    \be            {\begin{equation}}
\def    \ee            {\end{equation}}
\def    \bea           {\begin{eqnarray}}
\def    \eea           {\end{eqnarray}}

\def    \ler           {\stackrel{\scriptstyle <}{\scriptstyle\sim}} 
 
\begin{document}
%
\makeatletter
\@addtoreset{equation}{section}
\makeatother
\renewcommand{\theequation}{\thesection.\arabic{equation}}
\pagestyle{empty}
\rightline{FTUAM 99/11}
\rightline{IFT-UAM/CSIC-99-13}
\rightline{hep-ph/9904444}
\rightline{April 1999}
\vspace{0.0cm}
\begin{center}
\LARGE{Phenomenology of Non--Standard Embedding and Five--Branes
in M--Theory\\[10mm]}
\large{D.G. Cerde\~no and C. Mu\~noz\\[6mm]}
\small{
Departamento de F\'{\i}sica Te\'orica C-XI
and Instituto de F\'{\i}sica Te\'orica  C-XVI,\\[-0.3em]
Universidad Aut\'onoma de Madrid, 
Cantoblanco, 28049 Madrid, Spain. 
\\[7mm]}
\small{\bf Abstract} 
\\[7mm]
\end{center}
\begin{center}
\begin{minipage}[h]{14.0cm}

We study the phenomenology of the strong--coupling limit of
$E_8\times E_8$ heterotic string obtained from M--theory, 
using a Calabi--Yau compactification. After summarizing the
standard embedding results, we concentrate on 
non--standard embedding vacua as well as vacua where non--perturbative
objects as five--branes are present. We 
analyze in detail the different scales of the theory,
eleven--dimensional Planck mass, compactification scale, orbifold
scale,
and how they are related taking into account higher order corrections. 
To obtain the phenomenologically favored GUT scale 
is easier than in standard embedding vacua.
To lower 
this scale to intermediate ($\approx 10^{11}$ GeV) or $1$ TeV values  
or to obtain the radius of the orbifold as large as a millimetre
is possible. However, we point out that these special limits are unnatural.
Finally, we perform 
a systematic analysis of the soft supersymmetry--breaking terms. 
We point out that 
scalar masses larger than gaugino masses can easily be obtained 
unlike the standard embedding case and the weakly--coupled heterotic string.

\end{minipage}
\end{center}
\vspace{1.0cm}
\begin{center}
\begin{minipage}[h]{14.0cm}
PACS: 
04.65.+e, 12.60.Jv, 11.25.Mj, 11.25.-w

Keywords: 
scales, coupling unification, soft terms, Calabi--Yau compactification, 
heterotic string, M--theory
\end{minipage}
\end{center}
\newpage
\setcounter{page}{1}
\pagestyle{plain}
\renewcommand{\thefootnote}{\arabic{footnote}}
\setcounter{footnote}{0}
%
%
\section{Introduction}

One of the most exciting proposals of the last years in 
string theory, 
consists of the
possibility that the five distinct superstring theories in ten
dimensions plus
supergravity in eleven dimensions  
be different vacua in the moduli space of a single underlying 
eleven--dimensional 
theory, the so--called M--theory \cite{Schwarz}.
In this respect,
Ho\v{r}ava and Witten proposed that
the strong--coupling limit of  $E_8\times E_8$ heterotic
string theory can be obtained from M--theory. They used the low--energy
limit of M--theory, 
eleven-dimensional
supergravity, on  a manifold with boundary (a $S^1/Z_2$ orbifold), 
with the 
$E_8$ gauge multiplets at each of the 
two ten--dimensional boundaries (the orbifold fixed planes) 
\cite{Horava-Witten}.

Some phenomenological implications of the strong--coupling limit
of $E_8\times E_8$  heterotic string theory
have been studied by compactifying the eleven--dimensional M--theory 
on a Calabi--Yau manifold times
the eleventh segment (orbifold) \cite{Witten}.
The resulting four--dimensional effective theory can reconcile 
the observed Planck scale $M_{Planck}= 1.2 \times 10^{19}$ GeV 
with the phenomenologically favored GUT scale 
$M_{GUT}\approx 3\times 10^{16}$
GeV in a natural manner, providing an attractive framework
for the unification of couplings \cite{Witten,Banks-Dine}.
An additional phenomenological virtue of the M--theory limit
is that there can be a QCD axion whose high energy axion potential is
suppressed enough so that the  strong CP problem can be solved 
by the axion mechanism \cite{Banks-Dine,Choi}.
About the issue of supersymmetry 
breaking, the possibility of generating it by the
gaugino condensation on the hidden boundary has been studied
\cite{Horava,Nilles-Olechowski-Yamaguchi,Lalak-Thomas,Lukas-Ovrut-Waldram2,Quiros,Choi2}
and also some interesting features of the
resulting soft supersymmetry--breaking terms were discussed. 
In particular, gaugino masses turn
out to be of the same order as squark 
masses \cite{Nilles-Olechowski-Yamaguchi} unlike the 
weakly--coupled heterotic string case where gaugino masses are much smaller 
than squark masses \cite{Beatriz}. 
The analysis of the soft supersymmetry--breaking terms under the more general
assumption that supersymmetry is spontaneously broken by
the auxiliary components of the bulk moduli superfields in the model
(dilaton $S$ and modulus $T$) 
was carried out in \cite{Mio,Lukas-Ovrut-Waldram2}. 
It was examined in particular how the soft terms
vary when one moves from the weakly--coupled   heterotic string
limit to the strongly--coupled limit. The conclusion being 
that there can be a sizable difference between both limits.
As a consequence, the study
of the low--energy 
($\approx M_W$) sparticle spectra \cite{Mio,Lopez,Bailin,Kawamura,Savoy,Alejandro,Chi}
gives also rise to qualitative differences.

However, all the above mentioned analyses of the 
phenomenology of $N=1$ heterotic
M--theory vacua, were carried out only in the context of 
the standard embedding of the spin connection into one of the 
$E_8$ gauge groups. Although in the case of the weakly--coupled 
heterotic string is simple to work with the standard
embedding \cite{Candelas} and more involved the analyses of non--standard
embedding
vacua \cite{Distler,Kachru}, in the strongly--coupled case, as
emphasized in \cite{five-branes2}, the standard embedding is not
particularly special. 
Thus the analysis of the non--standard embedding 
is very interesting since more general gauge 
groups and matter fields may be present.   
Recently, this analysis has been considered 
in M--theory compactified on Calabi--Yau
\cite{Benakli,Lalak,five-branes,Benakli2} 
and orbifold \cite{Stieberger} spaces and several issues,
as the four--dimensional effective action and scales, studied. 

Concerning the latter 
it was pointed out in the past, in the context of perturbative
strings, 
that the size of the compactification scale might be of order $1$ TeV 
without any obvious contradiction with experimental facts
\cite{Antoniadis} (for a different point of view, see \cite{Caceres}).
Recently, going away from perturbative vacua, it was
realized that the 
string scale may be anywhere between the weak scale and the Planck
scale \cite{Lykken}. 
This is the case of the type I string where several interesting 
low string scale scenarios were proposed\footnote{             
To trust them would imply to assume that 
Nature is trying to mislead us with an apparent gauge coupling 
unification at the scale $M_{GUT}\approx 3\times 10^{16}$ GeV.
In this sense, a reasonable doubt about those scenarios is
healthy.}: a $1$ TeV scale scenario, 
where the size of the extra dimensions 
may even be as large as a millimetre \cite{Dimopoulos}
and an intermediate scale ($\approx 10^{11}$ GeV) 
scenario  where some phenomenological issues \cite{Benakli2} and the 
hierarchy $M_W/M_{Planck}\approx 10^{-16}$ \cite{Nosotros} can be explained. 
A $1$ TeV string scale scenario, with the scale of extra dimensions 
not smaller than $1$ TeV
may also arise in the context of type II strings \cite{Antoniadis2}.
Whether or not all these scenarios are possible in the context of M--theory
was analyzed recently in \cite{Benakli2} with interesting results.

On the other hand, more general vacua, still preserving $N=1$ supersymmetry,
may appear when $M5$--branes are included in the 
computation \cite{Witten}. 
These five--branes are non--perturbative objects, 
located at points throughout the orbifold interval.
They have $3+1$ uncompactified dimensions
in order to preserve Lorenz invariance and $2$ compactified dimensions.
The appearance of anomalous $U(1)$ symmetries related to their
presence
was studied in \cite{Binetruy}.
Modifications to the four--dimensional effective action
were discussed in the context of orbifold
compactifications of heterotic M--theory \cite{Stieberger} 
and investigated in great detail in Calabi--Yau compactifications 
by
Lukas, Ovrut and Waldram \cite{five-branes,five-branes2}. The latter also 
considered the 
issue of supersymmetry breaking by gaugino condensation and soft terms.
It is worth remarking that in the presence of five--branes,
model building turns out to be extremely interesting. 
In particular, to obtain three generation
models with realistic gauge groups, as for example $SU(5)$, is not
specially difficult \cite{Donagi}.

An important question in the study of string phenomenology is whether or 
not the soft terms, and in particular the scalar masses, are universal
\cite{Jerusalen}. In this respect, let us recall the
situation
in the case of Calabi--Yau compactifications of 
the weakly--coupled heterotic string. There the soft terms are in
general
non--universal due to the presence of off--diagonal matter K\"ahler
metrics
induced by the existence of different moduli ($T_i$) \cite{Hangbae}.
This can be avoided assuming that supersymmetry is broken in the
dilaton field ($S$) direction
\cite{Kaplunovsky-Louis,Brignole-Ibanez-Munoz2},
although {\it small} non--universality may arise due to string--loop
corrections \cite{Nir}\footnote{It is worth noticing
that supergravity--loop corrections may also 
induce non-universality \cite{Sik}.}.  

Concerning supersymmetry breaking of heterotic M--theory in a general
direction,
the situation
is similar to the weakly--coupled case: the existence in general
of different moduli, $T_i$, will give rise to non--universality. Moreover,
supersymmetry breaking in the dilaton direction will induce now 
{\it large} non--universal soft scalar masses due to the presence of
$S$ together with $T_i$ in the matter field K\"ahler metrics \cite{Mio}.
In \cite{five-branes2} an improvement  
to the problem of
non--universality in heterotic M--theory was
proposed: model building with Calabi--Yau spaces with only one K\"ahler
modulus $T$. Such spaces exist and, as mentioned above, 
in the presence of non--standard
embedding and five--branes, to construct three--generation models
might be relatively easy.

In this paper we will 
assume that the 
standard model arises from the heterotic M--theory 
compactified on a Calabi--Yau
manifold with only one field $T$, and then we will 
study the
associated phenomenology. 
In particular, we
will analyze first in detail the different scales of the theory,
eleven--dimensional
Planck mass, compactification scale, orbifold scale, and how they are 
related taking into account higher order corrections to the
formulae. In this respect, we will study whether or not large internal
dimensions are natural in the context of the heterotic M--theory.
We will also perform a systematic analysis of the soft
supersymmetry--breaking terms, under the general assumption that
supersymmetry is spontaneously broken by the bulk moduli fields.

In section 2 we will concentrate on
standard and non--standard embedding without the presence of
five--branes. First, we will summarize results about the standard
embedding
case concerning the four--dimensional effective action, which will be
very
useful for the detailed analysis of soft terms and scales of the
theory.
We will also 
compute the value of the $B$ parameter.
Then we will turn our thoughts to the study of the non--standard
embedding.
Basically, the same formulae than in the standard embedding case can be
used but with the value of one of the parameters which appears in 
the formulae in a different
range.
This will give rise to 
different possibilities for the scales of the theory and we will see
that to lower these scales is in principle possible in some
special limits. However, the necessity of a fine--tuning  
or the existence of a hierarchy problem renders this
possibility unnatural. Likewise, 
a different pattern of soft terms arises. For example, 
the possibility of scalar masses larger than gaugino masses,
something which is forbidden in the standard embedding case and
if possible, 
difficult to obtain, in the weakly--coupled heterotic string, is
now allowed.
In section 3 non--perturbative objects as five--branes are included in 
the vacuum. Thus the whole analysis is modified. New parameters 
contribute to the soft terms and their study is more involved. However,
an interesting pattern of soft terms arises. Again, scalars
heavier
than gauginos can be obtained but now more easily than in the non--standard 
embedding case. 
Concerning the scales, $M_{GUT}$ can easily be obtained. Extra
possibilities
to lower this scale arise but again they are unnatural.
Finally we leave the conclusions for section 4.

\section{Standard and non--standard embedding without five--branes}

\subsection{Four--dimensional effective action and scales}

Following Witten's investigation \cite{Witten}, the solution of the equations 
of motion, preserving N=1 supersymmetry, of eleven-dimensional M--theory
\cite{Horava-Witten} compactified on 
%
\be
M_4\times S^1/Z_2\times X\ ,
\label{space}
\ee
where $X$ is a six--dimensional
Calabi-Yau manifold and $M_4$ is four--dimensional
Minkowski space, can be analyzed by expanding it in powers of the
dimensionless parameter \cite{Banks-Dine}
\be
\epsilon_1 = \frac{\pi\rho}{M_{11}^{3} V^{2/3}}\ ,
\label{epsilon1}
\ee
where $M_{11}$ denotes the eleven-dimensional Planck mass, $V$ is the
Calabi-Yau volume and $\pi \rho$ denotes the length of the eleventh
segment. 

To zeroth order in this expansion, analogously to the case of
the
weakly--coupled heterotic string theory, two model--independent
bulk moduli superfields $S$ and $T$ arise in the four--dimensional
effective supergravity of the compactified M--theory. Their scalar
components can be identified as
\bea
S+\bar S &=&\frac{1}{\pi (4\pi)^{2/3}} M_{11}^6 V\ ,
\nonumber \\
T+\bar T &=& \frac{6^{1/3}}{(4\pi)^{4/3}}
 M_{11}^{3} V^{1/3} \pi\rho\ .
\label{dilaton-modulus}
\eea
Notice that with these definitions the expansion parameter
(\ref{epsilon1}) can be written as
\be
\epsilon_1 \approx 
 \frac{T+\bar T}{S+\bar S}\ .
\label{epsilon1bis}
\ee

To higher orders, there appear gauge and matter superfields associated
to the observable and hidden sector gauge groups, 
$G_O\times G_H \subset E_8\times E_8$, 
where $G_O$($G_H$) is located at the boundary 
$x^{11}=0$($x^{11}=\pi\rho$) with $x^{11}$ denoting the orbifold
coordinate.~From now on, we will use as our
notation the subscript $O$($H$) for quantities and functions of the
observable(hidden) sector. On the other hand, the internal space
becomes
deformed and is no longer as in (\ref{space}), a simple product of
$S^1/Z_2\times X$. 

The effective supergravity obtained from this
M--theory
compactification was computed to the leading order
in \cite{Banks-Dine,Li-Lopez-Nanopoulos1,Dudas-Grojean,Nilles-Olechowski-Yamaguchi}. The order $\epsilon_1$ correction to the leading order
gauge
kinetic functions and K\"ahler potential was also computed in 
\cite{Banks-Dine,Nilles-Stieberger,Choi,Nilles-Olechowski-Yamaguchi}
and
\cite{Lukas-Ovrut-Waldram} respectively. The final result 
is specified by the following K\"ahler potential, $K$,
gauge kinetic functions, $f_O$, $f_H$, and superpotential $W_O$:
\bea
K &=& -\ln (S+\bar{S}) -3\ln (T+\bar{T})
+\frac{3}{T+\bar{T}}
\left(1+\frac{1}{3}\epsilon_O\right) C_O^p \bar{C}_O^p 
\nonumber\\ &&
+\frac{3}{T+\bar{T}}
\left(1+\frac{1}{3}\epsilon_H\right) C_H^p \bar{C}_H^p \ ,
\label{kahler}
\\
f_{O} &=& S+\beta_O T\ , \quad f_{H}=S+\beta_H T\ , 
\label{kinetic}
\\
W_O &=& d_{pqr}C_O^pC_O^qC_O^r\ ,
\label{superpotential}
\eea
with
\be
\epsilon_O=\beta_O  
 \frac{T+\bar T}{S+\bar S}\ ,
\quad \epsilon_H=\beta_H  
 \frac{T+\bar T}{S+\bar S}\ .
\label{epsilonO}
\ee
$d_{pqr}$ are constant coefficients, $C_O^p(C_H^p)$ are the
observable(hidden)
matter fields and the model--dependent integer coefficients
$\beta_O={1\over 8\pi^2}\int\omega\wedge[{\rm tr}
(F_O\wedge F_O)-\frac{1}{2}{\rm tr}(R\wedge R)]$, 
$\beta_H={1\over 8\pi^2}\int\omega\wedge[{\rm tr}
(F_H\wedge F_H)-\frac{1}{2}{\rm tr}(R\wedge R)]$,
for the K\"ahler form $\omega$ normalized as
the generator of the integer (1,1) 
cohomology\footnote{
Usually $\beta$ is considered to be an arbitrary real number.
For $T$ normalized as (\protect\ref{dilaton-modulus}),
it is required to be an integer \cite{Choi}.}.

Taking into account that the real parts of the gauge kinetic functions 
in (\ref{kinetic}) multiplied by $4\pi$ are the inverse gauge coupling
constants $\alpha_O$ and $\alpha_H$, using (\ref{dilaton-modulus})
one can write \cite{Witten,Li-Lopez-Nanopoulos1,Benakli}
\bea
\alpha_O &=& \frac{1}{2\pi(S+\bar S)(1+\epsilon_O)}
\label{alphaO}
\\
&=& \frac{(4\pi)^{2/3}}{2 M_{11}^{6} V_O}\ ,
\label{alphaO'}
\eea
with 
%
%
\be
V_O=V(1+\epsilon_O)\ ,
\label{volumeO}
\ee
the observable--sector volume, and
\bea
\alpha_H &=& \frac{1}{2\pi(S+\bar S)(1+\epsilon_H)}
\label{alphaH}
\\ 
&=& \frac{(4\pi)^{2/3}}{2 M_{11}^{6} V_H}\ ,
\label{alphaH'}
\eea 
with 
%
%
\be
V_H=V(1+\epsilon_H)\ ,
\label{volumeH}
\ee
the hidden--sector volume.

On the other hand, using (\ref{volumeO}), 
the M-theory expression of the four--dimensional Planck scale 
%
\bea
M_{Planck}^2 &=& 16\pi^2\rho M_{11}^9 <V>\ ,
\label{Planck}
\eea
where $<V>$ is the average volume of the Calabi--Yau space
\bea
<V> &=& \frac{V_O+V_H}{2}
\label{averagevolume}
\eea
and (\ref{dilaton-modulus}) one also finds
\bea
V_O^{-1/6} &=& \left(\frac{V}{<V>}\right)^{1/2}
\left(\frac{6^{1/3}M_{Planck}^2}{2048 \pi^4}\right)^{1/2}
\left(\frac{4}{S+\bar S}\right)^{1/2}
\left(\frac{2}{T+ \bar T}\right)^{1/2} 
\left(\frac{1}{1+\epsilon_O}\right)^{1/6}
\nonumber\\
&=&
\left(\frac{V}{<V>}\right)^{1/2}
3.6\times 10^{16} 
\left(\frac{4}{S+\bar S}\right)^{1/2}
\left(\frac{2}{T+ \bar T}\right)^{1/2} 
\left(\frac{1}{1+\epsilon_O}\right)^{1/6}
{\rm GeV}
\ ,
\nonumber\\
\label{gut}
\eea
which is a very useful formula (also in the context of five--branes) 
as we will see below in order to 
discuss whether or not the GUT scale or smaller scales 
are obtained in a natural way. In this respect,
let us now obtain the connection between the different scales of the
theory:
the eleven--dimensional Planck mass, $M_{11}$, the
Calabi--Yau compactification scale, $V_O^{-1/6}$,
and the orbifold scale, $(\pi \rho)^{-1}$. 
It is straightforward to obtain from (\ref{alphaO'}) the following
relation:
\bea
\frac{M_{11}}{V_O^{-1/6}} &=& \left(2(4\pi)^{-2/3}\alpha_O\right)^{-1/6}\ .
\label{relation}
\eea
Likewise, using (\ref{Planck}) and (\ref{alphaO'}) 
we arrive at
\bea
\frac{V_O^{-1/6}}{(\pi \rho)^{-1}} &=& \left(\frac{V}{<V>}\right)
\left(\frac{M_{Planck}}{(2048\pi^4)^{1/2}V_O^{-1/6}}\right)^2 
\left(8192\pi^4\alpha_O^3\right)^{1/2}(1+\epsilon_O)
\nonumber\\
&=&
\left(\frac{V}{<V>}\right)
\left(\frac{2.7\times 10^{16}{\rm GeV}}{V_O^{-1/6}}\right)^2 
\left(8192\pi^4\alpha_O^3\right)^{1/2}(1+\epsilon_O)
\ .
\label{relation2}
\eea

\subsection{Standard embedding}

Here we will summarize results found in the literature about the
standard embedding case, including the computation of the 
soft terms under the general assumption of dilaton/modulus
supersymmetry
breaking,
and we will discuss in detail the issue of the scales 
in
the theory.

\begin{figure}[htb]
\begin{center}
\epsfig{file= 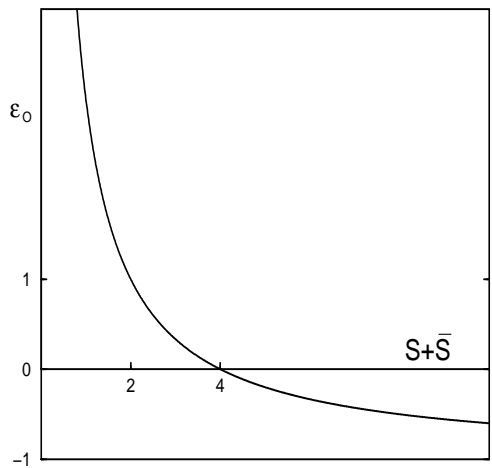, width=8cm, height=8cm}
\end{center} 
\vspace{-1.5cm}
\caption{$\epsilon_O$ versus $S+\bar S$.\label{epsilon}} 
\end{figure}

Let us recall first that in this case, where the spin connection of the
Calabi--Yau space is embedded in the gauge connection of one of the 
$E_8$ groups, the following constraint must be 
fulfilled:
\bea
\beta_O + \beta_H &=& 0\ ,
\label{constraint}
\eea
with\footnote{Non--standard embedding cases may also 
fulfil this condition. Since their effective action, as we will
discuss below, is the same as
in the standard embedding the results of this subsection
can be applied to any model obtained from standard or non--standard embedding 
with $\beta_O>0$. We leave the study of the case $\beta_O=0$ for the
next subsection.} 
\bea
\beta_O > 0\ . 
\label{mayorcero}
\eea
Thus $\epsilon_O=-\epsilon_H>0$
implying that the observable--sector volume $V_O$ given by  
(\ref{volumeO}) is larger than the hidden--sector volume $V_H$
given by (\ref{volumeH}) and therefore the gauge coupling of the observable
sector (\ref{alphaO'})
will be weaker than the gauge coupling of the hidden sector (\ref{alphaH'}). 
Besides, since $V_H$ must be a positive quantity, one has to impose the
bound $\epsilon_O<1$. Altogether one gets
\be
0 < \epsilon_O 
<
1\ .
\label{epsilonbound}
\ee
Notice that using (\ref{alphaO}) one can write
$\epsilon_O$ as
\bea
\epsilon_O 
=
\frac{4-(S+\bar S)}
{(S+\bar S)}\ ,
\label{nueva}
\eea
where we have already assumed that 
the gauge group of the observable sector
$G_O$ is the one of the standard model or some unification gauge group
as $SU(5)$, $SO(10)$ or $E_6$, i.e. we are using
$\left(2\pi\alpha_O\right)^{-1}=4$ in
order to reproduce the LEP data about 
$\alpha_{GUT}$ ($\alpha_O$ in our notation).
For a further study of scales and soft terms and comparison with
vacua in the presence of five--branes, we show $\epsilon_O$ versus $S+\bar S$
in Fig.~\ref{epsilon}.
The region between the lower bound $\epsilon_O=-1$ and $\epsilon_O=0$
corresponds to the non--standard embedding case and will be discussed
in subsection 2.3. The region with $\epsilon_O>1$ will be discussed
in the context of five--branes in section 3.
From (\ref{epsilonbound}), (\ref{nueva}) and (\ref{epsilonO})
one obtains that the dilaton and moduli fields are bounded (see also
Fig.~\ref{epsilon}) 
\be
0 < \beta_O(T+ \bar T) 
<
2\ , \quad 
2 
<
(S+ \bar S) 
<
4\ .
\label{bounds}
\ee
%
%
%
%
%
Finally, the average volume of the
Calabi--Yau space (\ref{averagevolume}) turns out to be equal to the
lowest order value
\bea
<V> &=& V\ ,
\label{constraint2}
\eea
and (\ref{gut}) simplifies as
\bea
V_O^{-1/6}
&=&
3.6\times 10^{16} 
\left(\frac{4}{S+\bar S}\right)^{1/2}
\left(\frac{2}{T+ \bar T}\right)^{1/2} 
\left(\frac{1}{1+\epsilon_O}\right)^{1/6}
{\rm GeV}\ ,
\label{gut2}
\eea
whereas (\ref{relation2})
becomes  
\be
\frac{V_O^{-1/6}}{(\pi \rho)^{-1}} =
7\ \left(\frac{2.7\times 10^{16}{\rm GeV}}{V_O^{-1/6}}\right)^2 
(1+\epsilon_O)
\ ,
\label{relation3}
\ee
where $\left(2\pi\alpha_O\right)^{-1}=4$ has been used.
This also allows us to write (\ref{relation}) as
\bea
\frac{M_{11}}{V_O^{-1/6}} &=& 2 \ .
\label{relationn}
\eea
%
%
%

With all these results we can start now the study of scales and soft
terms in the theory.

\subsubsection{Scales}

The four--dimensional effective theory from heterotic M--theory can
reconcile the observed $M_{Planck}=1.2\times 10^{19}$ GeV with
the phenomenologically favored GUT scale $M_{GUT}\approx 3\times 10^{16}$
GeV in a natural manner \cite{Witten,Banks-Dine}.
This is to be compared to the weakly--coupled heterotic string where
$M_{string}=\left(\frac{\alpha_{GUT}}{8}\right)^{1/2} M_{Planck} 
\approx 8.5\times 10^{17}$ GeV.
We will revisit this issue in detail taking into account the
higher order corrections studied above to the zeroth order formulae.

\begin{figure}[htb]
\begin{center}
\epsfig{file= 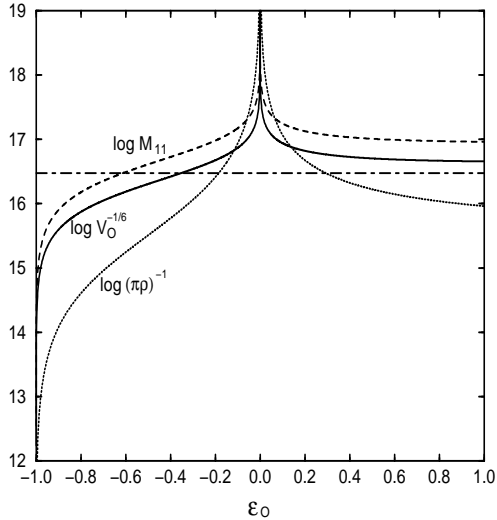, width=8cm, height=8cm}
\end{center} 
\vspace{-1.0cm}
\caption{$\log M_{11}$, $\log V_O^{-1/6}$ and $\log (\pi\rho)^{-1}$ 
versus $\epsilon_O$ 
in the cases $\beta_O=1$ (for $0<\epsilon_O<1$)
and $\beta_O=-1$ (for $-1<\epsilon_O<0$). The straight line
indicates the phenomenologically favored GUT scale,
$M_{GUT} = 3\times 10^{16}$ GeV.\label{scales1}} 
\end{figure}

Identifying $M_{GUT}\approx 3\times 10^{16}$ with $V_O^{-1/6}$
one obtains from (\ref{relation3}), with $\epsilon_O$ 
constrained by (\ref{epsilonbound}), and  (\ref{relationn}):
$M_{11}\approx 6\times 10^{16}$ GeV and
$(\pi\rho)^{-1}\approx (2.5-5.3)\times 10^{15}$ GeV, i.e.
the following pattern $(\pi\rho)^{-1}<V_O^{-1/6}<M_{11}$.
On the other hand, to obtain 
$V_O^{-1/6} \approx 3\times 10^{16}$ GeV when $\beta_O>0$ 
is quite natural. This can be seen from 
(\ref{gut2}) since (\ref{bounds}) implies that $T+\bar T$ and
$S+\bar S$ are essentially of order one.
Let us discuss this point in more detail.
Using (\ref{epsilonO}) and (\ref{nueva}) 
it is interesting to write (\ref{gut2}) as
\bea
V_O^{-1/6}
&=&
3.6\times 10^{16} 
\left(\frac{\beta_O}{2 \epsilon_O}\right)^{1/2}
\left(1+\epsilon_O\right)^{5/6}
{\rm GeV}\ .
\label{gut3}
\eea
This is shown in Fig.~\ref{scales1} where
$V_O^{-1/6}$ versus $\epsilon_O$ is plotted.
%
The right hand side of the figure ($0<\epsilon_O<1$)
corresponds to the case $\beta_O>0$ whereas the left hand side
($-1<\epsilon_O<0$) corresponds to the case  $\beta_O<0$, i.e.
the non--standard embedding situation that will be analyzed in the
next subsection.
For the moment we will concentrate on the case $\beta_O>0$ and,
in particular, in Fig.~\ref{scales1} we are showing an example with
$\beta_O=1$.
$(\pi\rho)^{-1}$ and 
$M_{11}$ are also plotted in the figure using (\ref{relation3}) and 
(\ref{relationn})
respectively. Most values of $\epsilon_O$ imply 
$V_O^{-1/6} \approx 5\times 10^{16}$ GeV which is quite close 
to the phenomenologically favored value. For example,
for $\epsilon_O=1/4$, which corresponds to $S+\bar S=16/5$ and 
$T+\bar T=4/5$, we obtain 
$V_O^{-1/6} = 6.1\times 10^{16}$ GeV
and for the limit\footnote{One may worry that the M--theory expansion would
not work in these cases where $\epsilon_O$ is of order one since 
$\epsilon_1$ in (\ref{epsilon1bis}) is of order one also. However,
as argued in \cite{Mio} any correction which is $n$-th order
in $\epsilon_1$ accompanies at least $(n-1)$-powers of 
$\epsilon_2=1/M_{11}^3 \pi\rho V^{1/3}\approx 1/2\pi^2(T+\bar T)$, 
the generalization of the
string world--sheet coupling to the membrane world--volume coupling,
and thus is suppressed by 
$\left(\epsilon_1\epsilon_2\right)^{n-1}\approx 
\left(\alpha_O/\pi\right)^{n-1}$. This allows the M--theory expansion
to be valid even when $\epsilon_1$ becomes of order one.
This has been explicitly checked in \cite{Wyllard}.}
$\epsilon_O=1$,
which corresponds to $S+\bar S=T+\bar T=2$, we obtain the lowest
possible
value  
$V_O^{-1/6} = 4.5\times 10^{16}$.

These qualitative results can only be modified in the limit 
$\epsilon_O\rightarrow 0$, i.e. $(T+\bar T)\rightarrow 0$,
since then $V_O^{-1/6}\rightarrow \infty$. Notice that in this case
$(\pi\rho)^{-1}>V_O^{-1/6}$ (see Fig.~\ref{scales1}).
This limit is not interesting not only because $V_O^{-1/6}$ is too
large but also because we are effectively in the weakly coupled 
region with a very small orbifold radius.

The results for $\beta_O\neq 1$ can easily be deduced from
the figure since 
$V_O^{-1/6}(\beta_O\neq 1)=\beta_O^{1/2} V_O^{-1/6}(\beta_O=1)$,
$M_{11}(\beta_O\neq 1)=2 V_O^{-1/6}(\beta_O\neq 1)$
and 
$(\pi\rho)^{-1}(\beta_O\neq 1)=\beta_O^{3/2}(\pi\rho)^{-1}(\beta_O=1)$.
Notice that for models with those values of $\beta_O$ we are in the limit of
validity if we want to obtain 
$V_O^{-1/6} \approx 3\times 10^{16}$ GeV.
For example, For $\epsilon_O=1$ with $\beta_O=4$, 
$V_O^{-1/6} = 9\times 10^{16}$ GeV.

Let us finally remark that, from the above discussion, it is 
straightforward to deduce that large internal dimensions,
associated with the radius of the Calabi--Yau and/or the radius of 
the orbifold, are not allowed.

\subsubsection{Soft terms}

Since the soft supersymmetry--breaking terms
depend on $\epsilon_O$ as we will see below, result 
(\ref{epsilonbound}) simplifies their
analysis. Applying the standard
(tree level) soft term formulae
\cite{Soni-Weldon,Brignole-Ibanez-Munoz} 
for the above supergravity model given by
(\ref{kahler}), (\ref{kinetic}) and (\ref{superpotential}), one can
compute the soft terms straightforwardly \cite{Mio}
\bea
M &=& \frac{\sqrt{3}Cm_{3/2}}{1+\epsilon_O}\left(\sin\theta e^{-i\gamma_S} + 
\frac{1}{\sqrt{3}}\epsilon_O\cos\theta e^{-i\gamma_T}\right)\ ,
\nonumber \\
m^2 &=& V_0 + m_{3/2}^2 
- 
\frac{3C^2m_{3/2}^2}{\left(3+\epsilon_O\right)^2}\left[
\epsilon_O\left(6+\epsilon_O\right)sin^2\theta + 
\left(3+2\epsilon_O\right)\cos^2\theta\right.     
\nonumber \\ &&
\left. 
- 2\sqrt{3}\epsilon_O\sin\theta\cos\theta \cos(\gamma_S-\gamma_T)\right]\ , 
\nonumber \\
A &=&
-\frac{\sqrt{3}Cm_{3/2}}{3+\epsilon_O}\left[\left(3-2\epsilon_O\right)
\sin\theta 
e^{-i\gamma_S} + \sqrt{3}\epsilon_O\cos\theta e^{-i\gamma_T} \right] \ .
\label{softterms}
\eea
Here $M$, $m$ and $A$ denote gaugino masses, scalar masses and 
trilinear
parameters respectively. The bilinear parameter, 
$B$, depends on the specific mechanism which could generate the
associated $\mu$ term \cite{Brignole-Ibanez-Munoz}. 
Assuming that the source of the $\mu$ term is a 
bilinear piece, $\mu (S,T)H_1H_2$, in the superpotential and/or
a bilinear piece, $Z(S,\bar S,T,\bar T)H_1H_2 +\ h.c.$, in the K\"ahler potential 
the result is
\bea
B &=& \hat \mu ^{-1} \left(\frac{T+\bar T}{3 + \epsilon_O}\right)
\left\{ \frac{\bar W(\bar S,\bar T)}{|W(S,T)|} 
(S+\bar S)^{-1/2}(T+\bar T)^{-3/2}\mu m_{3/2}C \left(-3\cos
\theta e^{-i\gamma_T}\right. \right.
\nonumber\\&&
-\sqrt{3}\sin\theta e^{-i\gamma_S}+
\frac{6\cos\theta e^{-i\gamma_T}}{3+\epsilon_O} 
+\frac{2\sqrt{3}\epsilon_O\sin\theta e^{-i\gamma_S}}{3+\epsilon_O}\nonumber 
-\frac{1}{C}
\nonumber\\&&
\left. +\frac{F^S}{m_{3/2}C}\partial_S\ln\mu+
\frac{F^T}{m_{3/2}C}\partial_T\ln\mu\right)
\nonumber \\ &&
+(2m_{3/2}^2+V_O)Z
-m_{3/2}^2C \left[ \sqrt{3}\sin\theta(S+\bar S)(\partial_{\bar S}Z 
e^{i\gamma_S}
-\partial_S Z e^{-i\gamma_S})\right.
\nonumber\\&&
\left. +\cos\theta(T+\bar T)(\partial_{\bar T}Z 
e^{i\gamma_T}-\partial_T Z e^{-i\gamma_T})\right]
\nonumber \\ &&
+ Zm_{3/2}^2C \left(
\frac{6\cos\theta e^{-i\gamma_T}}{3+\epsilon_O}
+\frac{2\sqrt{3}\epsilon_O\sin\theta e^{-i\gamma_S}}{3+\epsilon_O} \right)
\nonumber\\ &&
-m_{3/2}^2C^2\left[ 3\sin^2\theta (S+\bar S)^2 \partial_{\bar S}\partial_S Z
+ \cos^2\theta(T+\bar T)^2 \partial_{\bar T}\partial_T Z \right.
\nonumber\\ &&
+
\left(\frac{S+\bar S}{3+\epsilon_O}\right)
\left(6\epsilon_O\sin^2\theta\partial_{\bar S}Z +
6\cos^2\theta\frac{T+\bar T}{S+\bar S}\partial_{\bar T}Z\right)
\nonumber \\ &&
+\sqrt{3}\sin\theta\cos\theta \left(e^{i(\gamma_S-\gamma_T)}\partial_{\bar S}
\partial_T Z+ e^{-i(\gamma_S-\gamma_T)}\partial_{\bar T}\partial_S Z\right)
(S+\bar S)(T+ \bar T)
\nonumber \\ &&
+\frac{\sqrt{3}\sin\theta\cos\theta}{3+\epsilon_O}\left(  
6 e^{i(\gamma_S-\gamma_T)}(S+\bar S)\partial_{\bar S} Z \right.
\nonumber\\ &&
\left.\left.\left.
+
2\epsilon_O e^{-i(\gamma_S-\gamma_T)}(T+\bar T)\partial_{\bar T}
Z\right)
\right]
\right\}\ ,
\label{Bterm}
\eea
with the effective $\mu$ parameter given by:
\bea
\hat \mu &=& \left(\frac{T+\bar T}{3+ \epsilon_O}\right)
\left(\frac{\bar W(\bar S,\bar T)}{|W(S,T)|} 
(S+\bar S)^{-1/2}(T+\bar T)^{-3/2}
\mu \right.
\nonumber\\&&
\left. +m_{3/2}Z-\bar 
F^{\bar S}\partial _{\bar S}Z-\bar F^{\bar T}\partial _{\bar T}Z\right)
\ ,
\label{mu}
\eea
where $\partial_S(\partial_T)\equiv 
\frac{\partial}{\partial S}(\frac{\partial}{\partial T})$.
The part of (\ref{Bterm}) depending on $\mu$ was first computed in
\cite{Bailin,Kokorelis}.
We are using here the parameterization 
introduced in \cite{Brignole-Ibanez-Munoz2}
in order to know what fields, either $S$ or $T$, play the predominant role
in the process of SUSY breaking
\bea
F^S &=& \sqrt 3 m_{3/2}C(S+\bar S)\sin\theta e^{-i\gamma_S}\ ,
\nonumber \\
F^T &=& m_{3/2}C(T+\bar T)\cos\theta e^{-i\gamma_T}\ ,
\label{fterms}
\eea
with $m_{3/2}$ for the gravitino mass, $C^2=1+V_0/3m_{3/2}^2$ 
and $V_0$ for the (tree--level) vacuum energy density.

As mentioned in the introduction, the structure of these soft terms is
qualitatively different from
that of the weakly--coupled heterotic string
found in \cite{Brignole-Ibanez-Munoz2}, implying interesting 
low--energy ($\approx M_W$) phenomenology \cite{Mio,Bailin,Chi}.
In particular, in Fig.~1 of \cite{Mio} the dependence on 
$\theta$ of the soft terms for different values of $\epsilon_O$
in the range (\ref{epsilonbound}) is shown. For any value of
$\theta$, gauginos are heavier than scalars. We will come back to
discuss this point in more detail below.

\subsection{Non--standard embedding}

Although in the non--standard embedding case, there is no requirement
that the spin connection be embedded in the gauge connection, the
form of the effective action is still the same as in the standard--embedding
case, i.e. determined by 
(\ref{kahler}), (\ref{kinetic}) and (\ref{superpotential}).
Also constraint (\ref{constraint})
is still valid.
However, the relevant difference now is that the possibility
\bea
\beta_O < 0\ , 
\label{menorcero}
\eea
is allowed \cite{Benakli,Lalak,five-branes}. 
This is the case for example of two 
Calabi--Yau
models in \cite{Kachru}. One of them has $E_6$ as observable gauge
group
and three families with 
$\beta_O=-8$ \cite{Benakli}, and the other has  $SU(5)$ as observable
gauge group with $\beta_O=-4$ \cite{Bailin} .
We will revisit then the previous computations taking into account this
novel fact. In this sense we are concentrating in this subsection 
on non--standard embedding models with $\beta_O<0$. The study of those with
$\beta_O>0$ is included in the previous subsection.

Since $\epsilon_O=-\epsilon_H<0$, 
the volume $V_O$ in (\ref{volumeO}) is now smaller than
$V_H$ in (\ref{volumeH}) 
and therefore the
gauge coupling of the observable sector 
(\ref{alphaO'}) will be stronger than the one of the 
hidden sector\footnote{In the context of supersymmetry
breaking by gaugino condensation this scenario has several advantageous
features with respect to the standard embedding scenario. For a discussion
about this point see \cite{Lalak}.} (\ref{alphaH'}).
Besides, since $V_O$ must be a positive quantity, one has to impose the
bound $\epsilon_O>-1$. Altogether one gets
\be
-1 < \epsilon_O 
< 0\ ,
\label{epsilonbound'}
\ee
which corresponds, using (\ref{nueva}), to the bound 
(see also Fig.~\ref{epsilon}) 
\bea 
(S+\bar S)
> 4\ .
\label{bound1'}
\eea
%
%
%
%
%
%
%
Note that $\epsilon_O$ 
can approach the limit $-1$ only for very large values of 
($S+\bar S$) and therefore of ($T+\bar T$).

\subsubsection{Scales}

Let us now study how the scales are modified in these models with respect
to those  with $\beta_O>0$ 
studied in the previous subsection. We can use again
(\ref{gut3}), but now with $-1<\epsilon_O<0$.
This is shown in the left hand side 
of the Fig.~\ref{scales1}. 
Unlike the models of the previous subsection where always 
$V_O^{-1/6}$ was bigger than the GUT scale 
$3\times 10^{16}$ GeV for any $\beta_O>0$, in these non--standard
embedding models such a value can be obtained. For example in the
case shown in the figure, $\beta_O=-1$, with 
$\epsilon_O=-0.35$
which, using (\ref{nueva}) and (\ref{epsilonO}), 
corresponds to $S+\bar S=6.15$ and 
$T+\bar T=2.15$, we obtain
$V_O^{-1/6} = 3\times 10^{16}$ GeV.
For other values of $\beta_O$ this is also possible. Notice that, as
discussed in the previous subsection,  the
figure for $V_O^{-1/6}$ will be the same  adding the constant
$\log |\beta_O|^{1/2}$. So still there will be lines, corresponding
to $V_O^{-1/6}$, intersecting with the
straight
line corresponding to  
$M_{GUT} = 3\times 10^{16}$ GeV.
In this sense, if we want to obtain models with the
phenomenologically favored GUT scale, the non--standard embedding
is more compelling than cases with $\beta_O>0$.

On the other hand, 
in the previous subsection
we obtained the lower bound $10^{16}$ GeV for all scales of
the theory (see the right hand side of Fig.~\ref{scales1}),
far away from any direct experimental detection. Now we want
to study this issue in the non--standard embedding.
In fact, it was first pointed out in \cite{Benakli2}
that the possibility 
of lowering the scales of the theory with even 
an extra dimension as large as a millimetre
in some special limits is allowed in M--theory. We will analyze this
in detail clarifying whether or not such limits may be naturally
obtained.

~From (\ref{gut3}), clearly
in the limit $\epsilon_O\rightarrow -1$ we are able to obtain
$V_O^{-1/6}\rightarrow 0$ and therefore, given (\ref{relation3}) also
$(\pi\rho)^{-1}\rightarrow 0$ (see the left hand side of Fig.~\ref{scales1}).  
Thus to lower the scale $V_O^{-1/6}$ down to the experimental bound
(due to Kaluza--Klein excitations) of $1$ TeV 
is possible in this limit. However, this is true only
for values of $\epsilon_O$ extremely close to $-1$.
For example,
for $\epsilon_O=-0.999$ which, using (\ref{nueva}) and (\ref{epsilonO}), 
corresponds to $S+\bar S=4000$ and 
$T+\bar T=3996$, we obtain\footnote{Since
the unification scale has been lowered, the value
$(2\pi\alpha_O)^{-1}=4$ 
should be accordingly modified. However, the results are
not going to be essentially modified by this small change.}
$V_O^{-1/6} = 8\times 10^{13}$ GeV
and 
$(\pi\rho)^{-1} = 10^{11}$ GeV.
For $\epsilon_O=-0.999999$ corresponding to $S+\bar S=4\times 10^{6}$ and 
$T+\bar T=4\times 10^{6}-4$, we obtain the intermediate scale 
$V_O^{-1/6} = 2.5\times 10^{11}$ GeV, i.e. $M_{11}=5\times 10^{11}$ GeV,
with
$(\pi\rho)^{-1} = 3\times 10^{6}$ GeV. This is an interesting
possibility
since 
an intermediate scale $\approx 10^{11}$ GeV was proposed in
\cite{Benakli2} in order to solve some phenomenological problems and in
\cite{Nosotros} in order to solve the $M_W/M_{Planck}$ hierarchy.
In any case, 
it is obvious that the smaller the scale the larger the amount of 
fine--tuning becomes. The experimental lower bound for the scale $V_O^{-1/6}$,
$1$ TeV, can be obtained with
$\epsilon_O=10^{-16}-1$, i.e. 
$S+\bar S=4\times 10^{16}$ and 
$T+\bar T=4\times 10^{16}-4$. Then one gets
$V_O^{-1/6}=1181.5$ GeV with $(\pi\rho)^{-1} = 3.2\times 10^{-9}$ GeV.
Since only gravity is free to propagate in the orbifold, this extremely
small value is not a problem from the experimental point of view.
In any case,
it is clear that low scales are possible
but the fine--tuning needed renders the situation highly
unnatural. Another problem related with the limit 
$\epsilon_O\rightarrow -1$ will be found below when studying soft
terms,
since $|M|/m_{3/2}\rightarrow \infty$. Thus a extremely small gravitino
mass is needed to fine tune the gaugino mass $M$ to the $1$ TeV scale 
in order to avoid the gauge hierarchy problem.

\begin{figure}[htb]
\begin{center}
\epsfig{file= 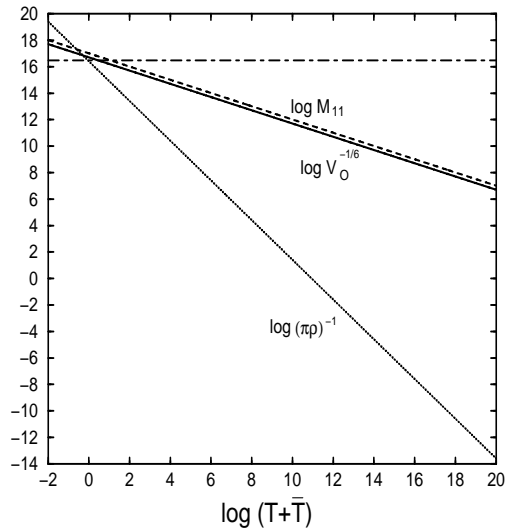, width=8cm, height=8cm}
\end{center} 
\vspace{-1.0cm}
\caption{$\log M_{11}$, $\log V_O^{-1/6}$ and $\log (\pi\rho)^{-1}$ 
versus $\log (T+\bar T)$ for the case $\beta_O=0$. 
The straight line
indicates the phenomenologically favored GUT scale,
$M_{GUT} = 3\times 10^{16}$ GeV.\label{scales2}} 
\end{figure}

There is a value of $\beta_O$ which is in principle allowed and has not been
analyzed yet. This is the case $\beta_O=0$. As we will see in a moment,
to lower the scales a lot in this context is again possible.
Since $\epsilon_O$ in (\ref{epsilonO}) is vanishing and using 
(\ref{nueva}),
$S+\bar S=4$, eq. (\ref{gut2}) can be written as 
\bea
V_O^{-1/6}
&=&
3.6\times 10^{16} 
\left(\frac{2}{T+ \bar T}\right)^{1/2} 
{\rm GeV}\ ,
\label{ggut2}
\eea
This is plotted in Fig.~\ref{scales2} together with 
$(\pi\rho)^{-1}$ and $M_{11}$. We see that the value 
$V_O^{-1/6} = 3\times 10^{16}$ GeV is obtained
for the reasonable value $T+\bar T=2.88$.


On the other hand, 
the larger $T+\bar T$ the smaller 
$V_O^{-1/6}$ becomes. In this way, 
for 
$T+\bar T=2.6\times 10^{11}$ GeV
one gets the intermediate scale $V_O^{-1/6}=10^{11}$ GeV, i.e.
$M_{11}=2\times 10^{11}$, with
$(\pi\rho)^{-1}=0.2$ GeV.
The lower bound for $V_O^{-1/6}$ is obtained 
with $T+\bar T=4\times 10^{19}$ GeV. Then
one gets $V_O^{-1/6}=8\times 10^{6}$ GeV and
$(\pi\rho)^{-1}=10^{-13}$ GeV. Smaller values of  
$V_O^{-1/6}$ are not allowed since experimental results on the
force of gravity constrain 
$(\pi\rho)$ to be less than a millimetre.
Thus, although low scales are allowed for the particular value
$\beta_O=0$, clearly a hierarchy problem between
$S+\bar S$ and $T+\bar T$ is introduced.

\subsubsection{Soft terms}

Since as mentioned above, the form of the effective action is still the
same as in the standard--embedding case,
in order to analyze the soft terms  
(\ref{softterms}) is still valid but for values of $\epsilon_O$ given
by (\ref{epsilonbound'}). This implies, as we will see below,
that the structure of these soft terms be qualitatively different
from that of the standard embedding case. 

\begin{figure}[htbp]
\hspace{-0.7cm}
\epsfig{file= 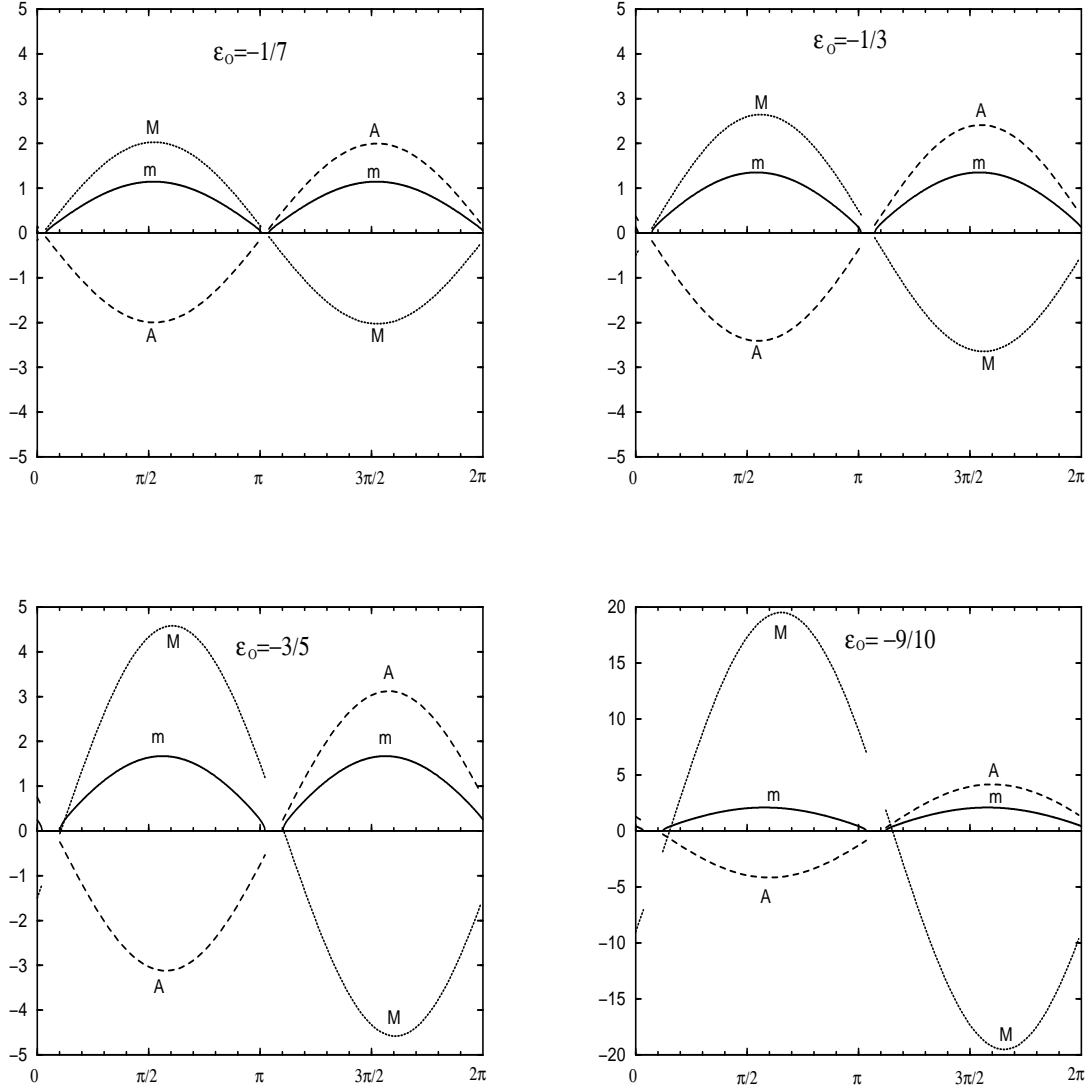, width=16cm, height=16cm}
\vspace{-1.0cm}
\caption{Soft parameters in units of $m_{3/2}$ versus $\theta$ for
different values of $\epsilon_O$ in the non--standard embedding case. Here
$M$, $m$
and $A$ are the gaugino mass, the scalar mass and the trilinear parameter
respectively.\label{nonstandardsoft}}
\end{figure}

In what follows, given the current experimental limits, 
we will assume $V_0=0$
and $\gamma_S=\gamma_T=0\ ({\rm mod}\ \, \pi)$. 
More specifically, we will set $\gamma_S$ and $\gamma_T$ 
to zero and allow $\theta$ to vary in a range $[0, 2\pi)$. 
We show in Fig.~\ref{nonstandardsoft} the dependence on $\theta$
of
the soft terms $M$, $m$, and $A$ in units of the gravitino mass for
different values of $\epsilon_O$.
Notice that for 
$\theta \in [\pi, 2\pi)$ the corresponding figures could have easily been 
deduced
since the shift $\theta\rightarrow \theta + \pi$ in (\ref{softterms})
implies $m\rightarrow m$, $M\rightarrow -M$ and $A\rightarrow -A$.
However we prefer to plot them explicitly to compare with the case
which will be 
studied in the next section 
where due to the inclusion of five--branes 
this
symmetry is broken (see e.g. Fig.~\ref{susyfivesoft}). 
Several comments are in order.
First of all, some (small) ranges of $\theta$ are forbidden
by having a  negative scalar mass-squared.
About the possible range of soft terms,
the smaller the value of $\epsilon_O$, the larger the range becomes.
For example, 
for 
$\epsilon_O=-1/3$, 
those ranges are
$0.1<|M|/m_{3/2}<2.65$, $0<m/m_{3/2}<1.35$ and $0.18<|A|/m_{3/2}<2.41$,
whereas
for $\epsilon_O=-3/5$, 
they are
$0<|M|/m_{3/2}<4.58$, $0<m/m_{3/2}<1.67$ and $0.25<|A|/m_{3/2}<3.12$.


This result is to be compared with the one of \cite{Mio} where the
standard-embedding soft terms were analyzed. From Fig.~1 of that paper,
where cases $\epsilon_O=1/7,1/3,3/5,1$ where shown, we see that the 
forbidden ranges of $\theta$ are now substantially decreased
For example, whereas the dilaton--dominated case, $|\sin\theta| = 1$
is always allowed, in the standard--embedding situation it may be
forbidden
depending on the value of $\epsilon_O$. Also we see that the range
of soft terms is now substantially increased. In the limit
$\epsilon_O\rightarrow -1$, $0.3<|A|/m_{3/2}<4.58$,  
$0<m/m_{3/2}<2.26$ and $|M| \rightarrow \infty$.

\begin{figure}[htbp]
\begin{center}
\epsfig{file= 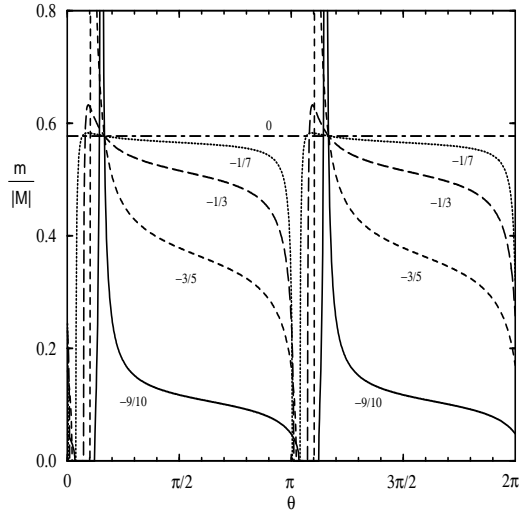, width=8cm, height=8cm}
\end{center}
\vspace{-1.5cm}
\caption{$m/|M|$ versus $\theta$ for different values of $\epsilon_O$ 
in the non--standard embedding case.\label{nonstandardcociente}}
\end{figure}

In order to discuss the SUSY spectra further, it is worth
noticing that although gaugino masses are in general larger than 
scalar masses, for values of $\epsilon_O$ approaching $-1$ there are two  
narrow ranges of values of $\theta$ where the opposite situation
occurs. 
This can be seen in Fig.~\ref{nonstandardsoft} 
for the cases $\epsilon_O=-3/5, -9/10$.
Let us remark that $M/m_{3/2}$ and $m/m_{3/2}$ are then very
small
and therefore $m_{3/2}$ must be large in order to fulfil e.g. the
low--energy
bounds on gluino masses \footnote{This is a similar situation to 
that of the weakly--coupled orbifold scenario (O-II) of 
\cite{Brignole-Ibanez-Munoz2}. There for untwisted particles, in the limit 
$\sin\theta\rightarrow 0$, scalar and gaugino masses vanish at tree
level:
then string loop effects become important and tend to make scalars 
heavier than gauginos.}.
These ranges of $\theta$ can be seen in more detail
in Fig.~\ref{nonstandardcociente}, 
where the ratio $m/|M|$ versus $\theta$ is plotted 
for different values of $\epsilon_O$.
This result is to be compared with the one of Fig.~3 in
\cite{Mio}.
There, the standard embedding case is analyzed and 
$r\equiv m/|M|< 1$
for any value of
$\theta$.
Fig.~\ref{nonstandardcociente} also allows us to study some properties of the low--energy 
($\approx M_W$)
spectra
independently of the details of the electroweak breaking
using the formula (see e.g. \cite{Mio})
\bea
&& 
\hspace{-1.0cm}
M_{\tilde g}:m_{\tilde Q_L}:m_{\tilde u_R}:m_{\tilde d_R}:m_{\tilde L_L}:
m_{\tilde e_R} 
\nonumber \\
&& 
\hspace{-1.0cm}
\approx 
1:\frac{1}{3}\sqrt {7.6+r^2}:\frac{1}{3}\sqrt {7.17+r^2}:
\frac{1}{3}\sqrt {7.14+r^2}:\frac{1}{3}\sqrt{0.53+r^2}:\frac{1}{3}
\sqrt{0.15+r^2}\ ,
\label{masas}
\eea
where $\tilde g$ denote the gluino, 
$\tilde l$ all the sleptons and $\tilde q$
first and second generation squarks. Most values of $\theta$ imply
$r<1$ and from (\ref{masas}) we obtain
\bea
M_{\tilde g}\approx m_{\tilde q}>m_{\tilde l}\ . 
\label{gluinos1}
\eea
This is also the generic (tree--level) result in Calabi--Yau compactifications 
of the weakly--coupled heterotic
string, which can be recovered from (\ref{softterms}) by taking the
limit
$(T+\bar T)<<(S+\bar S)$, i.e. $\epsilon_O\rightarrow 0$.
Then $M=\sqrt {3} m=\sqrt {3} m_{3/2} \sin\theta$ \cite{Brignole-Ibanez-Munoz2}
implying $r=1/\sqrt 3$ except in the limit $\sin\theta\rightarrow 0$
which is not well defined. 
Only in this limit one might expect $r>1$,
similarly to what happens in the
orbifold case, due to string loop corrections. This is something to be 
checked \cite{NUEVA}.  
      
However, when $\epsilon_O$ 
is approaching $-1$ the above situation (\ref{gluinos1}) 
concerning Fig.~\ref{nonstandardcociente} can be reversed since
\bea
M_{\tilde g}<m_{\tilde q}\approx m_{\tilde l}\ , 
\label{gluinos2}
\eea
for 
$\theta$ in the narrow ranges where $r>1$.

Finally, notice that in the case $\beta_O=0$, i.e. $\epsilon_O=0$, 
which corresponds to the straight line $m/|M|=1/\sqrt 3$, 
the phenomenologically interesting sum--rule 
$\sum_{i=1}^3 m_i^2=M^2$ \cite{Scheich}
is trivially fulfilled. Moreover, it 
is also fulfilled for any other possible value 
\footnote{We thank T. Kobayashi for putting forward this fact to us.} 
of $\epsilon_O$
when $\theta=\pi/6$ (and $\pi/6 + \pi$) since then the 
$\epsilon_O$ contribution in (\ref{softterms}) is cancelled and one 
obtains $m=m_{3/2}/2$ and
$|M|=\sqrt 3 m_{3/2}/2$.
See also in Fig.~\ref{nonstandardcociente} 
how all graphs intersect each other at those angles.
Since this result is independent on the value of $\epsilon_O$, it
happened also for $0<\epsilon_O<1$ (see Fig.~3 in \cite{Mio}).
%
%


\section{Vacua with five-branes}

In the previous section, we studied the phenomenology of heterotic
M--theory vacua obtained through standard and non--standard
embeddings.
Here we want to analyze (non--perturbative) heterotic M--theory vacua
due to the presence of five--branes \cite{Witten}. 
As mentioned in the introduction, five--branes are non--perturbative objects, 
located at points, $x^{11}=x_n (n=1,...,N)$, throughout the orbifold interval.
The modifications to the four--dimensional effective action determined by
(\ref{kahler}), (\ref{kinetic}) and (\ref{superpotential}), 
due to their presence, have recently been investigated by
Lukas, Ovrut and Waldram \cite{five-branes,five-branes2}. Basically,
they are due to existence of moduli, $Z_n$, whose
${\rm Re}(Z_n)\equiv z_n=x_n/\pi\rho \in (0,1)$ 
are the five--brane positions in the normalized orbifold coordinates.
Then, the effective supergravity obtained from heterotic M--theory
compactified 
on a Calabi--Yau manifold in the presence of five--branes is now determined by
\bea
K &=& -\ln (S+\bar{S}) -3\ln (T+\bar{T})+ K_5 + \frac{3}{T+\bar{T}}
\left(1+\frac{1}{3}
e_O\right) H_{pq}C_O^p \bar{C}_O^q,
\nonumber\\
f_{O} &=& S+
B_O
T\ , \quad f_{H}=S+
B_H
T\ ,\nonumber \\
W_O &=& d_{pqr}C_O^pC_O^qC_O^r\ ,
\label{kahlerbranas}
\eea
with
\be
e_O=b_O  
 \frac{T+\bar T}{S+\bar S}\ .
\label{Es}
\ee
Here
$K_5$ is the K\"ahler potential for the five--brane moduli
$Z_n$, $H_{pq}$ is some $T$--independent metric and
\bea
b_O &=& \beta_O + \sum_{n=1}^N(1-z_n)^2\beta_n\ ,
\nonumber \\
B_O &=& \beta_O + \sum_{n=1}^N(1-Z_n)^2\beta_n\ ,
\nonumber \\
B_H &=& \beta_H + \sum_{n=1}^N(Z_n)^2\beta_n\ ,
\label{Bes}                   
\eea
with $\beta_O$, $\beta_H$ the instanton numbers and $\beta_n$
the five--brane charges. The former, instead of constraint (\ref{constraint}),
must fulfil the following constraint:
\be
\beta_O + \sum_{n=1}^N\beta_n+\beta_H=0\ .
\label{Besbis}
\ee

It is worth noticing that in addition to the observable and hidden
sector
gauge groups, $G_O$ and $G_H$, there appear gauge groups 
from the five--branes. Thus the total gauge group at low energies is in fact 
$G_O\times G_H\times G_1\times ...\times G_N$ \cite{five-branes,five-branes2}. 
We are assuming here
that $G_O$ arises from one of the $E_8$
groups. In this sense the five--brane sectors are considered
hidden sector interacting with the observable sector only through
bulk supergravity.

Let us also remark that we are considering, as in the previous section,
a compactification on
a Calabi--Yau manifold with only one K\"ahler modulus
$h_{1,1}=1$. As discussed in the introduction, such Calabi--Yau spaces
exist and their phenomenological properties are extremely interesting.

Assuming for simplicity that $<Z_n>=<z_n>$, i.e. $<B_O>=<b_O>$, the set
of eqs. (\ref{alphaO})--(\ref{volumeH}) is still valid with the
modification
$\epsilon_{O,H}\rightarrow e_{O,H}$, where
\bea
e_H=b_H\frac{T+\bar T}{S+\bar S}\ , 
\label{lala}
\eea
with 
\bea
b_H = \beta_H + \sum_{n=1}^N(z_n)^2\beta_n\ . 
\label{Besbis2}
\eea
Following the
analysis
of subsection 2.2 
%
%
we can write $e_O$ 
as
\bea
e_O 
=
\frac{4-(S+\bar S)}
{(S+\bar S)}\ ,
\label{nueva2}
\eea
%
and therefore
Fig.~\ref{epsilon} is still valid substituying $\epsilon_O$ by $e_O$.
%
%
We can obtain different bounds on $e_O$ 
depending
on the sign of both $b_O$ and $b_H$:

\noindent {\it i)} $b_H\geq 0$, $b_O\leq 0$

Then  $e_H$ is positive and $e_O$ negative. Since
$V_O=V(1+e_O)$ must be positive we need
%
\bea
-1<e_O\leq 0\ .
\label{cota1}
\eea

\noindent {\it ii)} $b_H\geq 0$, $b_O>0$

Now since $e_O$ is positive $V_O$ will always be positive and 
therefore
the only bound is 
%
\bea
0<e_O\ .
\label{cota2}
\eea

\noindent {\it iii)} $b_H<0$, $b_O>0$

In this case $e_H$ is negative. Since
$V_H=V(1+e_H)$ must be positive we need
$e_H>-1$. On the other hand, using (\ref{Es}) and (\ref{lala}),
$e_O=e_H\frac{b_O}{b_H}$ 
and as a consequence the following bounds are obtained
%
\bea
0<e_O<\frac{b_O}{|b_H|}\ .
\label{cota3}
\eea

\noindent {\it iv)} $b_H<0$, $b_O\leq 0$

Now both $e_O$ and $e_H$ are negative. To avoid negative volumes 
we need $e_O>-1$ and $e_H>-1$. As discussed above we can write
$e_O=e_H\frac{b_O}{b_H}$ and therefore two possibilities arise:

If $|b_O|\geq |b_H|$
%
\bea
-1<e_O\leq 0\ .
\label{cota5}
\eea

If $|b_O|<|b_H|$ 
%
\bea
-\frac{b_O}{b_H}< e_O\leq 0\ .
\label{cota4}
\eea
For instance, for the example studied in \cite{five-branes}
%
%
where there are four five--branes at
$(z_1,z_2,z_3,z_4)=(0.2,0.6,0.8,0.8)$
with charges $(\beta_1,\beta_2,\beta_3,\beta_4)=(1,1,1,1)$
and instanton number $(\beta_O,\beta_H)=(-1,-3)$, one obtains
$b_O=-0.12$, $b_H=-1.32$ and then one should apply 
(\ref{cota4}) with the result $-0.09\ler e_O\ler 0$.

The 
standard and non--standard
embeddings studied in section 2 are particular cases of this more
general
analysis with five--branes. 
For $\beta_n=0$ in (\ref{Bes}) and (\ref{Besbis}), i.e. no five--branes, 
we have $b_{O}=\beta_{O}=-\beta_H=-b_H$ 
and $e_O=\epsilon_O$. If $\beta_O<0$,
from {\it i)}  
we recover the non--standard embedding case, $-1<\epsilon_O<0$. This is the
part of the graph corresponding to $S+\bar S>4$ shown in Fig.~\ref{epsilon}.
For  $\beta_O>0$, from {\it iii)} we recover the standard embedding 
(and also some non--standard
embedding) case since  $0<\epsilon_O<1$ 
This is the part of the graph corresponding to $2<S+\bar S<4$ shown in 
Fig.~\ref{epsilon}.
It is worth noticing that the values $0<(S+\bar S)<2$, corresponding to 
$\epsilon_O>1$,
which are not possible in the absence of five--branes,
are allowed in their presence since 
$e_O>1$ is possible (cases {\it ii)}
and {\it iii)}).

\subsection{Scales}

\begin{figure}[htb]
\hspace{-0.7cm}
\epsfig{file= 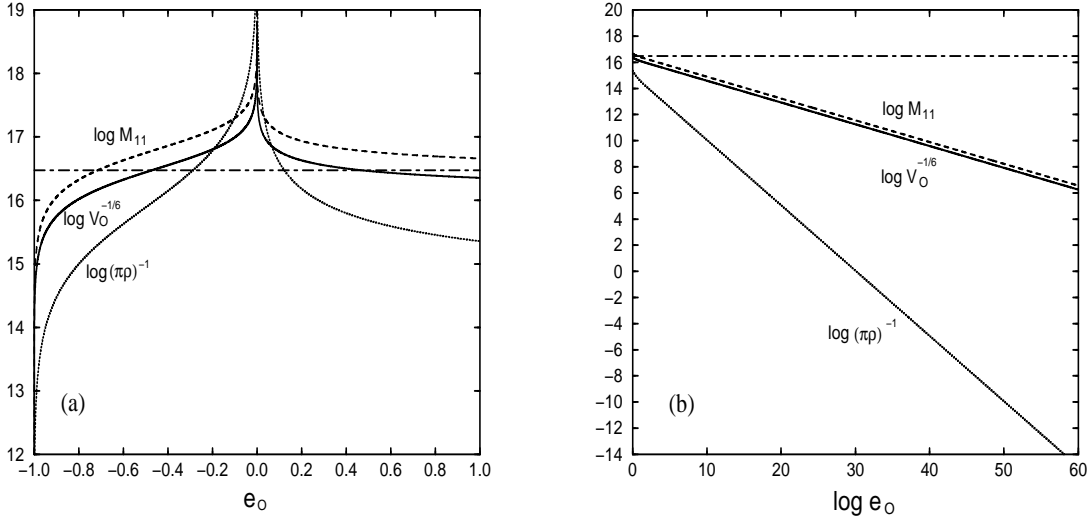, width=16cm, height=8cm}
\vspace{-1.0cm}
\caption{$\log M_{11}$, $\log V_O^{-1/6}$ and $\log (\pi\rho)^{-1}$ 
versus (a) $e_O$ and (b) $\log e_O$ 
in the cases $b_O=b_H=1/2$ (for $0<e_O$)
and $b_O=-7/4$, $b_H=5/4$ (for $-1<e_O<0$). The straight line
indicates the phenomenologically favored GUT scale,
$M_{GUT} = 3\times 10^{16}$ GeV.\label{scales1five}} 
\end{figure}

In the presence of five--branes (\ref{constraint2}) is no longer true since
(\ref{volumeO}) and (\ref{volumeH}) are modified in the following way:
$V_O=V(1+e_O)$ and $V_H=V(1+e_H)$. Therefore
\be
<V>=V \left( 1+ \frac{e_O+e_H}{2}\right)
\ .
\label{medio}
\ee
Then the relevant formulae to study the relation between the different
scales of the theory are (\ref{gut})  and
(\ref{relation2}) with the modification $\epsilon_O\rightarrow e_O$.
Notice that (\ref{relation}) is not modified.
Similarly to the case without five--branes, to obtain
$V_O^{-1/6}\approx 3\times 10^{16}$ GeV when 
$T+\bar T$ and $S+\bar S$ 
are of order one is quite natural. This can be seen
from (\ref{gut})
\bea
V_O^{-1/6} &=& 
\left(\frac{1}{1+\frac{e_O}{2}\left(1+\frac{b_H}{b_O}\right)}\right)^{1/2} 
3.6\times 10^{16}
\left(\frac{4}{S+\bar S}\right)^{1/2}
\left(\frac{2}{T+ \bar T}\right)^{1/2} 
\left(\frac{1}{1+e_O}\right)^{1/6}
{\rm GeV} 
\nonumber\\
\label{puf}
\eea
where (\ref{medio}) with (\ref{lala}) and (\ref{Es}) has been used.
Using (\ref{Es}) and (\ref{nueva2}) it is interesting to write 
(\ref{puf}) as
\bea
V_O^{-1/6}
&=&
\left(\frac{1}{1+\frac{e_O}{2}\left(1+\frac{b_H}{b_O}\right)}\right)^{1/2} 
3.6\times 10^{16} 
\left(\frac{b_O}{2 e_O}\right)^{1/2}
\left(1+e_O\right)^{5/6}
{\rm GeV}\ .
\label{niidea}
\eea
In the left hand side of the Fig.~\ref{scales1five}a
we show an example of the case {\it i)}, $b_O=-7/4$ and $b_H=5/4$,
corresponding to  $-1<e_O<0$. In the right hand side 
we show an example of the case {\it ii)}, $b_O=b_H=1/2$,
corresponding to $e_0>0$. Both are interesting examples since 
they cover the whole range of validity of $e_O$ and will be used below
to study the soft terms. Unlike the standard and non--standard
embedding
cases with $\beta_O>0$ shown in the right hand side of Fig.~\ref{scales1},
now the line corresponding to  $V_O^{-1/6}$ intersects easily
the straight line corresponding to the GUT scale. This is obtained 
for $e_O=0.46$ which, using 
(\ref{nueva2}) and (\ref{Es}), 
corresponds to $S+\bar S=2.73$ and 
$T+\bar T=2.54$.
Of course this effect is due essentially 
to the extra factor appearing in (\ref{niidea}), coming from the
average
volume,
with respect to (\ref{gut3}).
%

Only in some special limits one may lower the scales. 
As in the case without fivebranes, fine--tuning
$e_O\rightarrow -1$
we are able to obtain  $V_O^{-1/6}$ as low as we wish.
The numerical results will be basically similar 
to the ones of non--standard embedding in subsection 2.3.1

Moreover, $e_O>1$ is possible in the presence of five--branes.
Therefore with $e_O$ sufficiently large we may get $V_O^{-1/6}$ 
very small. This is shown in Fig.~\ref{scales1five}b.
For example, with $\log e_O=56.1$ the experimental lower bound 
$(\pi\rho)^{-1}=10^{-13}$ GeV
is obtained for  $V_O^{-1/6}=8\times 10^{6}$ GeV, corresponding  
to $S+\bar S=3.1\times 10^{-56}$ and 
$T+\bar T=8$.
Clearly we introduce a hierarchy problem.

\begin{figure}[htb]
\begin{center}
\epsfig{file= 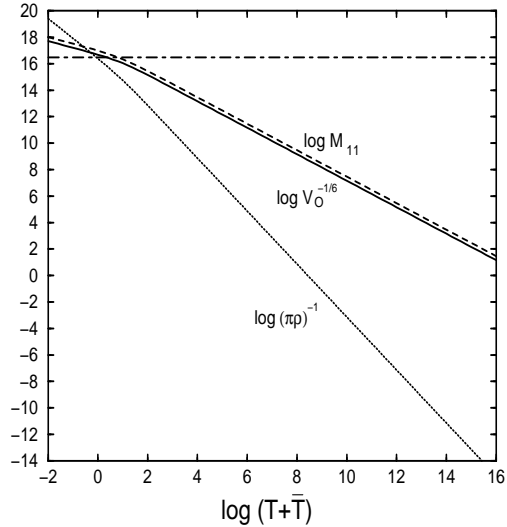, width=8cm, height=8cm}
\end{center} 
\vspace{-1.0cm}
\caption{$\log M_{11}$, $\log V_O^{-1/6}$ and $\log (\pi\rho)^{-1}$ 
versus $\log (T+\bar T)$ for the case $b_O=0$. 
The straight line
indicates the phenomenologically favored GUT scale,
$M_{GUT} = 3\times 10^{16}$ GeV.\label{scales2five}} 
\end{figure}

Finally, 
Let us analyze the special case $b_O=0$. The analysis will be
similar
to the one of the case $\beta_O=0$ without five--branes in subsection
2.3.1.
Since $e_O$ in (\ref{Es}) is vanishing and using (\ref{nueva2}),
$S+\bar S=4$, eq. (\ref{gut}) can be written as 
\bea
V_O^{-1/6}
&=&
\left(\frac{1}{1+\frac{1}{8}b_H(T+\bar T)}\right)^{1/2} 
3.6\times 10^{16} 
\left(\frac{2}{T+ \bar T}\right)^{1/2} 
{\rm GeV}\ ,
\label{media}
\eea
where (\ref{lala}) has been used. Depending on the value of $b_H$
we obtain different results. If $b_H=0$, eq. (\ref{ggut2}) is recovered
and therefore we obtain the same results as in the case
$\beta_O=\beta_H=0$ without five--branes (see Fig.~\ref{scales2}).
If $b_H>0$ the results are qualitatively similar, the larger
$T+\bar T$ the smaller $V_O^{-1/6}$ becomes. However, notice that now
for large $T$ 
we have a factor $(T+\bar T)^{-1}$ and then not so large values of 
$T+\bar T$  as in Fig.~\ref{scales2} are needed in order to lower the scales.
This is shown in Fig.~\ref{scales2five} for the case $b_H=1$.
Then $V_O^{-1/6}=1$ TeV can be obtained for $T+\bar T=10^{14}$ 
with the size of
the orbifold $(\pi\rho)^{-1}=5\times 10^{-12}$ GeV close to  its experimental
bound of $1$ millimetre. 
In any case, still a large hierarchy between $S+\bar S$ and
$T+\bar T$ is needed.
Finally, for $b_H<0$ we are in the case  {\it iv)} and therefore we
have 
the constraint 
$0<(T+\bar T)<4/|b_H|$, implying that $V_O^{-1/6}$ around the GUT scale
can be obtained but lowering it to intermediate or $1$ TeV values is
not
possible.

\subsection{Soft terms}

Let us now concentrate on the computation of soft terms. The above 
supergravity model (\ref{kahlerbranas}) gives rise to the following 
gaugino masses, scalar masses and trilinear parameters:
\bea
M &=& \frac{1}{(S+\bar S)(1+\frac{B_O T + \bar B_O \bar T}{S+ \bar S})} 
\left(F^S + F^T B_O + T F^n {\partial_n} {B_O}  \right)\ , 
\nonumber \\ 
m^2 &=& V_0 + m_{3/2}^2 -\frac{1}{(3+e_O)^2}
\left[e_O(6+e_O) \frac{|F^S|^2}{(S+\bar S)^2} \right.
\nonumber\\ &&
+3(3+2e_O)
\frac{|F^T|^2}{(T+\bar T)^2}  
-\frac{6e_O}{(S+\bar S)(T+\bar T)} \preal F^S \bar F^{\bar T} 
\nonumber\\&&
+ \left(\frac{e_O}{b_O}(3+e_O)
\partial_n\partial_{\bar m} b_O - \frac{e_O^2}{b_O^2}
\partial_n b_O \partial_{\bar m} b_O \right) F^n\bar F^{\bar m}
\nonumber \\ &&
\left. - 
\frac{6e_O}{b_O}\frac{\partial_{\bar n} b_O }{S+\bar S}
\preal F^S \bar F^{\bar n}
+ \frac{6e_O}{b_O}\frac{\partial_{\bar n} b_O}{T+\bar T}
\preal 
F^T \bar F^{\bar n}
\right]\ ,
\nonumber\\
A&=&-\frac{1}{3+\epsilon_O}\left[
(3-2e_O)\frac{F^S}{S+\bar S}+ 
3e_O \frac{F^T}{T+\bar T} \right. 
\nonumber\\ &&
+  \left. \left( \frac{3e_O}{b_O}\partial_n b_O
-(3+ e_O){\partial _n} K_5
\right) F^n \right]\ , 
\label{softfive}
\eea
where $\partial_n\equiv \frac{\partial}{\partial Z_n}$ and for
the moment we write explicitly the $F$--terms of the dilaton ($F^S$),
modulus ($F^T$) and five--branes ($F^n$). They must fulfil the relation
\be
V_0=\frac{|F^S|^2}{(S+\bar S)^2}+\frac{3|F^T|^2}{(T+\bar T)^2}+
\bar F^{\bar n} F^m 
\partial_{\bar n} \partial_m K_5 -3m_{3/2}^2 \ .
\label{potencial}
\ee
Assuming, as in subsection~2.2.2, that the source of the $\mu$ term
is a bilinear piece in the superpotential and/or the K\"ahler
potential,
the result is 
\bea
B&=& \hat\mu^{-1}\left(\frac{T+\bar T}{3+e_O}\right)\left\{\frac{\bar W(\bar S,\bar T)}{|W(S,T)|}\ (S+\bar S)^{-1/2}(T+\bar T)^{-3/2}e^{K_5/2}\mu\right.
\nonumber \\ 
&&\left[ F^S\left(\frac{-1}{S+\bar S}+\right.\right.
\left.\partial_S\log\mu+\frac{2e_O}{3+e_O}\frac{1}{S+\bar S}\right)
+F^T\left(\frac{-3}{T+\bar T}+\right.\partial_T\log\mu
\nonumber \\ 
&&\left.+\frac{6}{3+e_O}\frac{1}{T+\bar T}\right)+
F^n\left(\part_nK_5+\partial_n\log\mu-\frac{2\part_nb_O}{3+e_O}\right.
\left.\frac{T+\bar T}{S+\bar S}\right)
-\left.m_{3/2}\right]
\nonumber\\
&&+
(2m_{3/2}^2+V_0)Z-m_{3/2}
\left(\bar F^{\bar S}\partial_{\bar S}Z+\right.
\left.\bar F^{\bar T}\partial_{\bar T}Z+\bar F^{\bar n}\part_{\bar n}Z\right)
\nonumber\\
&&+m_{3/2}
\left[F^S\left(\partial_SZ+\frac{2e_O}{3+e_O}\right.\right.
\left.\frac{1}{S+\bar S}Z\right)+
F^T\left(\partial_TZ+\frac{6}{3+e_O}\right.
\left.\frac{1}{T+\bar T}Z\right)
\nonumber\\
&&
+F^n\left(\partial_nZ-\frac{2\part_nb_O}{3+e_O}\right.
\left.\left.\frac{T+\bar T}{S+\bar S}Z\right)\right]-
|F^S|^2
\left(\part_S\part_{\bar S}Z+\frac{2e_O}{3+e_O}\frac{\part_{\bar S}Z}{S+\bar S}\right)
\nonumber\\
&&-
|F^T|^2
\left(\part_T\part_{\bar T}Z+\frac{6}{3+e_O}\frac{\part_{\bar T}Z}{T+\bar T}\right)
-\bar F^{\bar n}F^m
\left(\part_m\part_{\bar n}Z-\frac{2\part_mb_O}{3+e_O}\right.
\left.\frac{T+\bar T}{S+\bar S}\part_{\bar n}Z\right)
\nonumber\\
&&-\bar F^{\bar S}F^T
\left(\part_T\part_{\bar S}Z+\frac{6}{3+e_O}\frac{\part_{\bar S}Z}{T+\bar T}\right)-
\bar F^{\bar T}F^S
\left(\part_S\part_{\bar T}Z+\frac{2e_O}{3+e_O}\frac{\part_{\bar T}Z}{S+\bar S}\right)
\nonumber\\
&&-\bar F^{\bar S}F^n
\left(\part_n\part_{\bar S}Z-\frac{2\part_nb_O}{3+e_O}\right.
\left.\frac{T+\bar T}{S+\bar S}\part_{\bar S}Z\right)-
\bar F^{\bar n}F^S
\left(\part_S\part_{\bar n}Z+\frac{2e_O}{3+e_O}\frac{\part_{\bar n}Z}{S+\bar S}\right)
\nonumber\\
&&-\bar F^{\bar T}F^n
\left(\part_n\part_{\bar T}Z-\frac{2\part_nb_O}{3+e_O}\right.
\left.\frac{T+\bar T}{S+\bar S}\part_{\bar T}Z\right)-
\bar F^{\bar n}F^T
\left.\left(\part_T\part_{\bar n}Z+\frac{6}{3+e_O}\frac{\part_{\bar
n}Z}{T+\bar T}\right)\right\}\ ,
\label{bzeta}
\nonumber \\
\eea
with the effective $\mu$ parameter given by:
\bea
\hat\mu&=&\left(\frac{T+\bar T}{3+e_O}\right)
\left(\frac{\bar W(\bar S,\bar T)}{|W(S,T)|}(S+\bar S)^{-1/2}(T+\bar T)^{-3/2}e^{K_5/2}\mu\right.
+ m_{3/2}Z
\nonumber \\ 
&&\left.-\bar{F}^{\bar S}\part_{\bar S}Z-\bar{F}^{\bar T}\part_{\bar T}Z-\bar{F}^{\bar n}\part_{\bar n}Z\right)
\eea

\begin{figure}[htb]
\hspace{-0.7cm}
\epsfig{file= 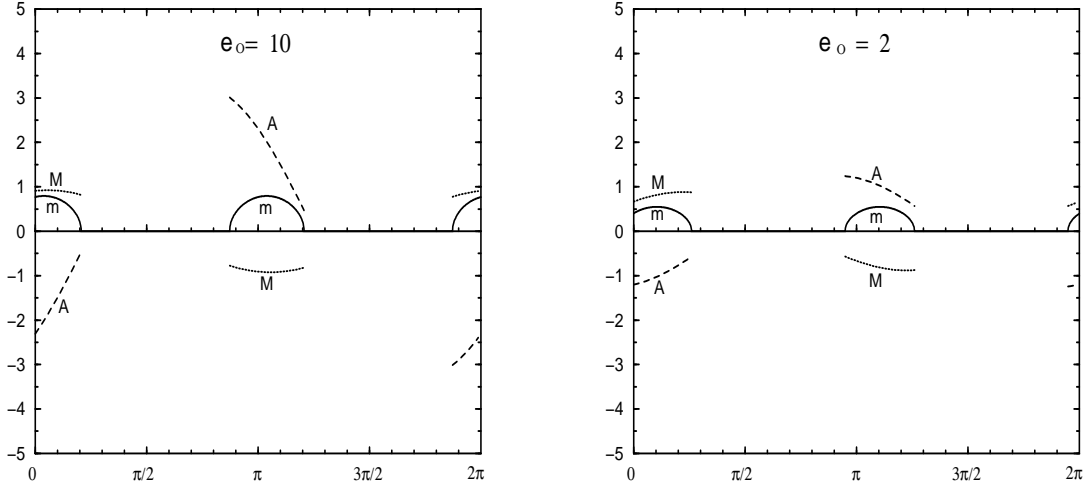, width=16cm, height=8cm} 
\vspace{-1.0cm}
\caption{Soft parameters in units of $m_{3/2}$ versus
$\theta$ for 
different values of $e_O$ when five--branes are present without
contributing to supersymmetry breaking. 
\label{nonsusyfivesoft}}
\end{figure}

\begin{figure}[htb]
\begin{center}
\epsfig{file= 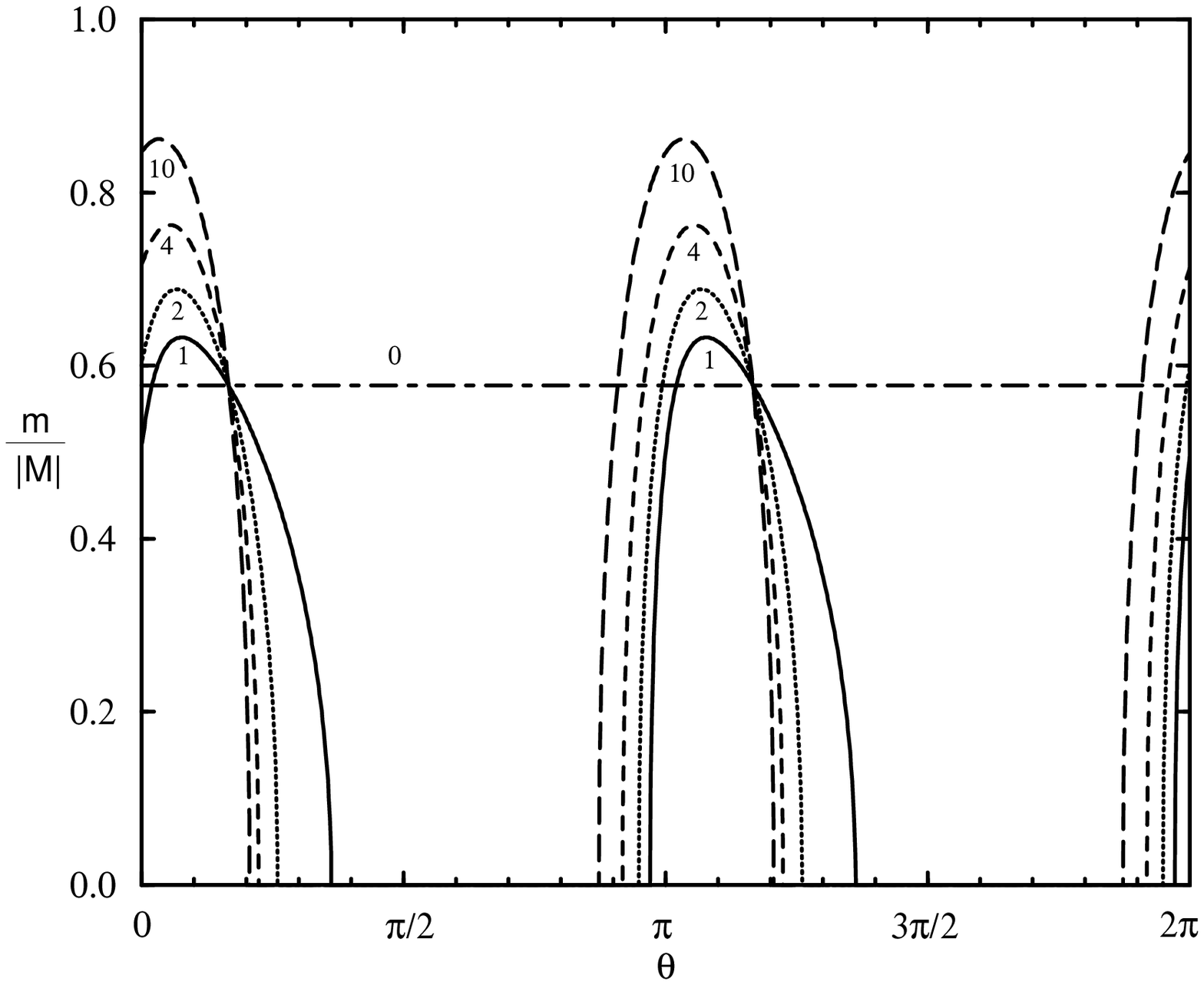, width=8cm, height=8cm}
\end{center}  
\vspace{-1.5cm}
\caption{$m/|M|$ versus $\theta$ 
for different values of $e_O$
when five--branes are present without contributing to supersymmetry
breaking.\label{nonsusyfivecociente}}
\end{figure}

Due to the possible contribution of several $F$--terms associated with 
five--branes, which can have in principle off--diagonal
K\"ahler
metrics, the numerical computation of the soft terms turns out to be
extremely
involved. In order to get an idea of their value and also to study
the deviations with respect to the case without five--branes we can
do some simplifications. One possibility is to assume that 
five--branes are present but only the
$F$--terms
associated with the dilaton and the modulus contribute to supersymmetry
breaking,
i.e. $F^n=0$. Then, assuming as before $<Z_n>=<z_n>$
and using parametrization (\ref{fterms}), eq. (\ref{softfive}) reduces to
eq. (\ref{softterms})
with $e_O$ instead of $\epsilon_O$.
Under these simplifying assumptions, Figs.~\ref{nonstandardsoft} 
and \ref{nonstandardcociente} in section 2
(and Figs. 1 and 3 in \cite{Mio}) are also valid in this case 
since, as discussed above 
the
range of allowed values of $e_O$ includes those of $\epsilon_O$, 
i.e. $-1<e_O<1$. The relevant difference with respect to the case
without
five--branes is that now values with $e_O\geq 1$ are allowed.
We
plot this possibility
in Figs.~\ref{nonsusyfivesoft} and \ref{nonsusyfivecociente} 
for some values of $e_O$. 
Fig.~\ref{nonsusyfivesoft} shows that the range of $\theta$ forbidden
by having a negative scalar mass--squared, is large and quite stable
against variations of $e_O$. This
pattern of soft terms is qualitatively different from that without
five--branes analyzed in subsection 2.3 for the non--standard embedding
and in Fig.~1 of \cite{Mio} for the standard embedding.
However, the fact that  always scalar  masses are smaller than gaugino masses
in the latter case (see Fig.~3 in \cite{Mio}) is also true here. In 
Fig.~\ref{nonsusyfivecociente} 
we show the ratio $m/|M|$ versus $\theta$ for different values
of $e_O$. Although the larger the value of
$e_O$, the larger the ratio becomes, there is an upper bound 
$m/|M|=1$ (obtained for $\theta =0$). As discussed in (\ref{gluinos1})
we will obtain at low--energies,
$M_{\tilde g}\approx m_{\tilde q}>m_{\tilde l}$.
Notice that the sum--rule $3m^2=M^2$ is also fulfilled for 
$\theta=\pi/6$ (and $\pi/6 + \pi$).
%
%
%
%

Another possibility to simplify the numerical computation of the soft 
terms is to assume that there is only one five--brane in the model. 
For example,
parametrizing consistently with (\ref{potencial})
\bea
F^S &=& \sqrt{3} m_{3/2} C (S+\bar S)\sin\theta\cos\theta_1 
e^{-i\gamma_S} \ ,
\nonumber\\
F^T &=&  m_{3/2} C (T+\bar T)\cos\theta\cos\theta_1 e^{-i\gamma_T}\ ,
\nonumber \\
F^1 &=& \sqrt{3}  m_{3/2} C 
({{\partial}_1{\partial}_{\bar 1 }K_5})^{-1/2}sin\theta_1 
e^{-i\gamma_1}\ ,
\label{ftermss}
\eea
where $\theta_1$ is the new goldstino angle associated to the $F$--term
of the five--brane $F^1$, we obtain from (\ref{softfive})
\bea
M&=&\frac{ \sqrt{3} m_{3/2}C}
{\left(1+\frac{B_O T + \bar B_O \bar T}{S+ \bar S}\right)}
\left( \sin\theta\cos\theta_1
e^{-i\gamma_S}+\frac{1}{\sqrt{3}}B_O  \frac{e_O}{b_O} 
\cos\theta\cos\theta_1 
e^{-i\gamma_T} \right.
\nonumber \\&&
\left. -\frac{2T}{S+\bar S}(1-Z_1)\beta_1 
({{\partial}_1{\partial}_{\bar 1 }K_5})^{-1/2}
sin\theta_1 e^{-i\gamma_1} \right)  \ ,
\nonumber \\
m^2&=&V_o + m_{3/2}^2-\frac{3 m_{3/2}^2C^2}{(3+e_O)^2}\left\{ e_O 
(6+e_O)\sin^2\theta 
\cos^2\theta_1 \right.
\nonumber \\&&
+(3+2e_O)\cos^2\theta\cos^2\theta_1  
- 2\sqrt{3}e_O\sin\theta\cos\theta\cos^2\theta_1\cos(\gamma_S 
-\gamma_T) 
\nonumber \\&&
+ ({{\partial}_1{\partial}_{\bar 1 }K_5})^{-1}
\sin^2\theta_1 \left( (3+e_O)  
\beta_1\frac{e_O}{2b_O} -\left[(1-z_1) 
\beta_1 \frac{e_O}{b_O}\right]^2 \right)
\nonumber \\&&
+6(1-z_1)\beta_1\frac{e_O}{b_O}
({{\partial}_1{\partial}_{\bar 1 }K_5})^{-1/2}
\sin\theta\sin\theta_1\cos\theta_1\cos(\gamma_1
-\gamma_S)
\nonumber \\&&
\left. -2\sqrt{3}(1-z_1)\beta_1\frac{e_O}{b_O} 
({{\partial}_1{\partial}_{\bar 1 }K_5})^{-1/2}
\cos\theta\sin\theta_1\cos\theta_1\cos(\gamma_T
-\gamma_n)\right\}\ , 
\nonumber\\
A&=&-\frac{\sqrt{3} m_{3/2}C}{3+e_O}\left[ (3-2e_O) 
\sin\theta\cos\theta_1 e^{-i\gamma_S} + \sqrt{3}e_O\cos\theta\cos\theta_1
e^{-i\gamma_T} \right.
\nonumber \\&&
\left. -
({{\partial}_1{\partial}_{\bar 1 }K_5})^{-1/2}
sin\theta_1 e^{-i\gamma_1} 
\left((3+e_O)
{{\partial}_1 K_5}
 + 3(1-z_1)\beta_1 \frac{e_O}{b_O} \right) \right] \ .
\label{una}
\eea
%
%
%
\begin{figure}[htb]
\hspace{-0.7cm}
\epsfig{file= 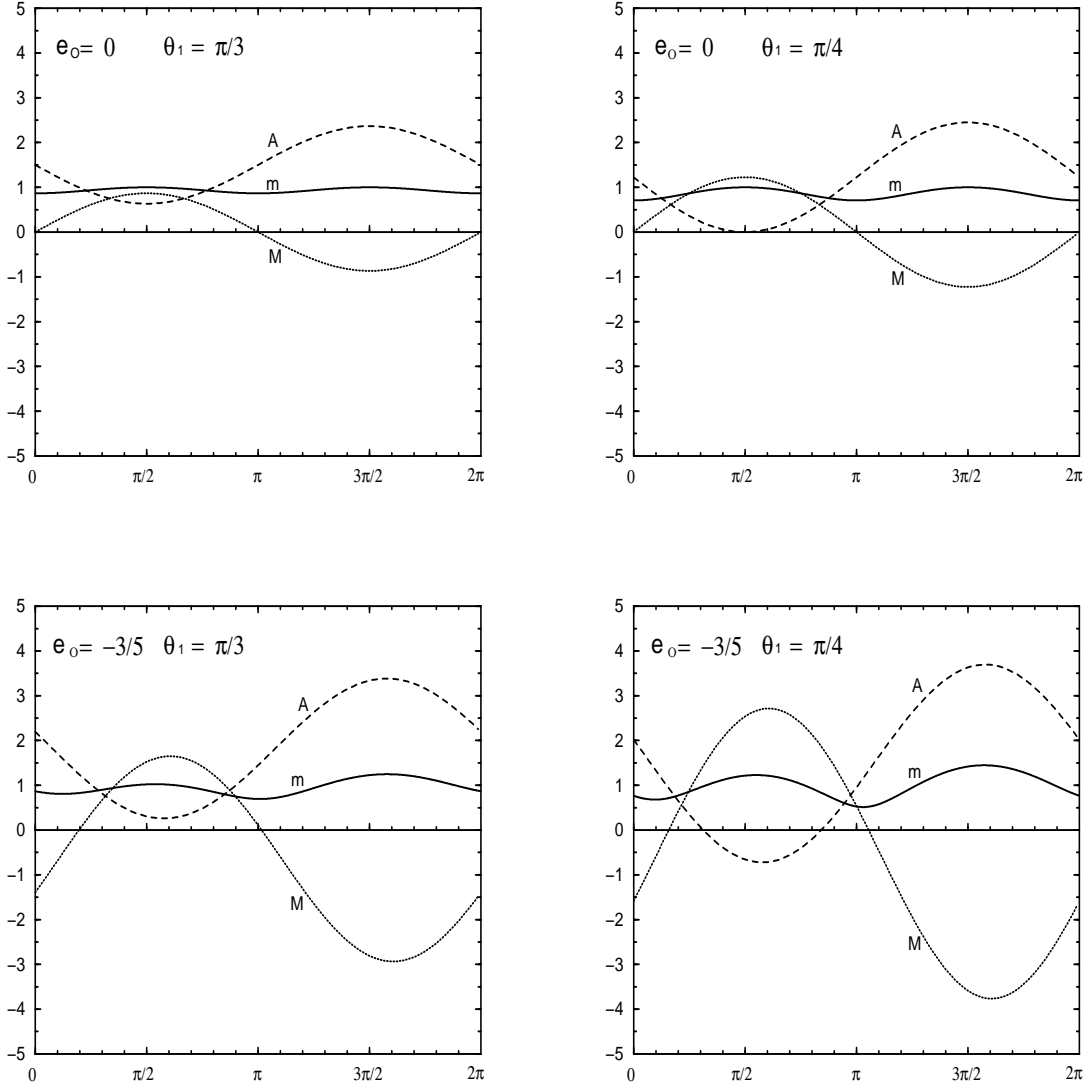, width=16cm, height=16cm}
\vspace{-1.0cm}
\caption{Soft parameters in units of $m_{3/2}$ versus $\theta$ for
different values of $e_O$ and 
goldstino angle $\theta_1$
when one five--brane is
contributing to supersymmetry breaking.\label{susyfivesoft}}
\end{figure}
Similarly one can obtain the $B$ parameter using (\ref{bzeta}).

Unfortunately, the numerical analysis of this simplified case is not
straightforward. All soft terms depend not only on the new goldstino
angle $\theta_1$ in addition to 
$m_{3/2}$, $\theta$
 and $e_O$, but also on other free parameters.
For example, although gaugino masses can be further 
simplified with the assumption
$<Z_n>=<z_n>$, i.e. $<B_O>=<\bar B_O>=<b_O>$, and $<T>=<\bar T>$
\bea
M&=&\frac{ \sqrt{3} m_{3/2}C}
{1+e_O}
\lbrace \sin\theta\cos\theta_1 e^{-i\gamma_S}
+\frac{1}{\sqrt{3}} e_O 
\cos\theta\cos\theta_1 e^{-i\gamma_T}
\nonumber \\&&
-\frac{e_O}{b_O}(1-z_1)\beta_1 
({{\partial}_1{\partial}_{\bar 1 }K_5})^{-1/2}
sin\theta_1 e^{-i\gamma_1}    \rbrace\ , 
\label{gauginos}
\eea
%
still
they have an explicit dependence on 
$z_1$ and 
$\partial_1 \partial_{\bar 1} K_5$.
Notice that, for a given model,
$\beta_O$ and $\beta_1$ are known and therefore $b_O$ can be computed
from (\ref{Bes}) once $z_1$ is fixed.
Something similar occurs for the 
$A$ parameter, where $z_1$, $\partial_1 K_5$ and
$\partial_1 \partial_{\bar 1} K_5$ appear explicitly, and
for the scalar masses, where 
 $z_1$ and
$\partial_1 \partial_{\bar 1} K_5$ also appear.
Thus in order to compute soft terms when a five--brane is present and
contributing to supersymmetry breaking we have to input these values.
Fortunately, $z_1$ is in the range $(0,1)$ and, although
$K_5$ is not known,
since
it depends on $z_1$,
we expect 
$\partial_1 K_5$, $\partial_1 \partial_{\bar 1} K_5 = {\cal O}(1)$.
So we can consider the following representative example:
$z_1=1/2$ and $\partial_1 K_5=\partial_1\partial_{\bar 1}K_5=1$.
Since, still we have to input the value of $b_O$, we choose the
interesting
example with $\beta_O=-2$ and $\beta_1=1$ which implies $b_O=-7/4$.
In this way, using (\ref{Besbis}) and (\ref{Besbis2}), $b_H$ is also fixed,
$b_H=5/4$, and from the result (\ref{cota1}) in case {\it i)} we know that the
whole range of allowed (negative) values of $e_O$ can be
analyzed.

%
%
\begin{figure}[htb]
\hspace{-0.7cm}
\epsfig{file= 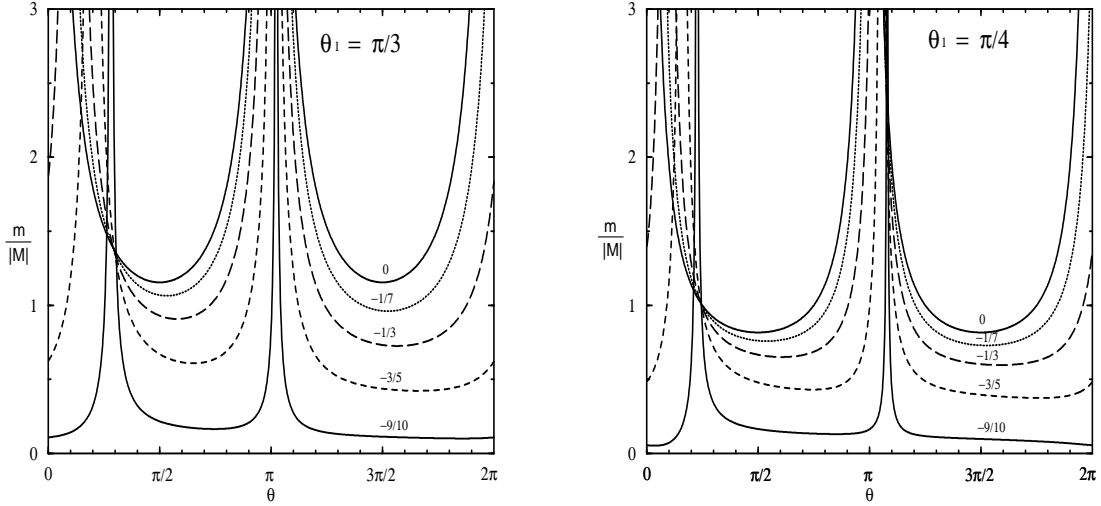, width=16cm, height=8cm}
\vspace{-1.0cm}
\caption{$m/|M|$ versus $\theta$ for different values of 
$e_O$ and 
goldstino angle $\theta_1$
when one five--brane is
contributing to supersymmetry breaking.\label{susyfivecociente}}
\end{figure}
%
%
\begin{figure}[htb]
\hspace{-0.7cm} 
\epsfig{file= 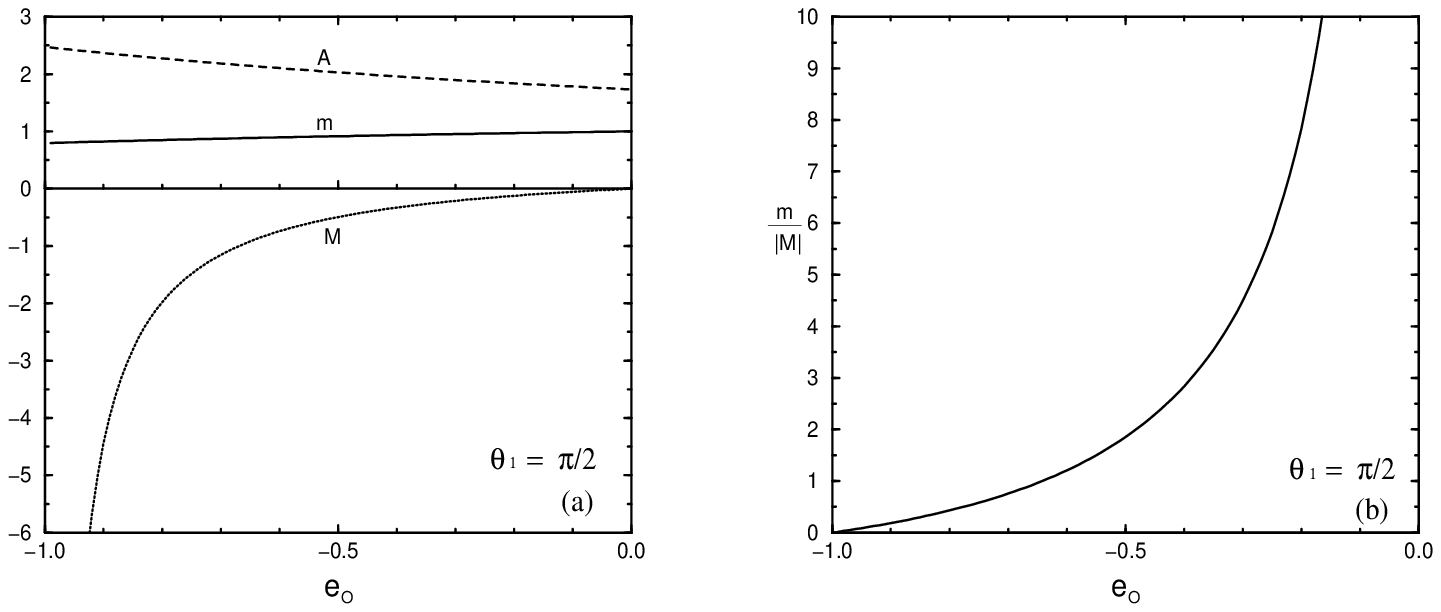, width=16cm, height=8cm}
\vspace{-1.0cm}
\caption{(a) Soft parameters in units of $m_{3/2}$ and (b)
$m/|M|$ versus $e_O$, 
when only the five--brane contributes to supersymmetry breaking with
$\theta_1=\pi/2$.\label{susyonlyfivesoft}}
\end{figure}

Taking
into account again the current experimental limits, we will assume
$V_0=0$
and 
$\gamma_S=\gamma_T=\gamma_1=0\ ({\rm mod}\ \, \pi)$. More specifically
we will set $\gamma_S$, $\gamma_T$ and $\gamma_1$ to zero and allow
$\theta$ and $\theta_1$ to vary in a range $[0,2\pi)$. We show in 
Fig.~\ref{susyfivesoft} 
the dependence on $\theta$ of the soft terms in units of the 
gravitino mass for some values of $e_O$ and $\theta_1$. 
Due to the contribution of 
$\theta_1$ to the soft terms (\ref{una}) the shift 
$\theta\rightarrow \theta + \pi$ does not imply, as in 
Figs.~\ref{nonstandardsoft} and \ref{nonsusyfivesoft},
$m\rightarrow m$, $M\rightarrow -M$ and $A\rightarrow -A$.
However, this is still true for $\theta_1\rightarrow \theta_1 + \pi$.
Besides, under the shifts $\theta_1\rightarrow \pi - \theta_1$
and $\theta\rightarrow \theta + \pi$, $m$, $M$ and $A$ remain
invariant.
Therefore from the analysis of the figures corresponding to 
$\theta_1\in [0,\pi/2]$ the rest of the figures can easily be deduced.
In particular, in Fig.~\ref{susyfivesoft} we plot the values of $\theta_1$,
$\pi/4$ and $\pi/3$. 
For a fixed
value of $\theta_1$, as in the case
without five--branes shown in
Fig.~\ref{nonstandardsoft} (with $\epsilon_O$ instead of $e_O$),  
the smaller the value of $e_O$ the larger the range of soft terms becomes.
However,
unlike Fig.~\ref{nonstandardsoft}, 
in Fig.~\ref{susyfivesoft} we see a remarkable fact: scalar masses
larger
than gaugino masses can easily be obtained. This happens not only for narrow
ranges
of $\theta$ but even for the whole range (see the figure
with $e_O=0$ and $\theta_1=\pi/3$). We can see this in more detail in
Fig.~\ref{susyfivecociente} where $m/|M|$ versus $\theta$ is plotted.

For example, $m/|M|>1$ 
can be obtained, for any value of $\theta$, for values  
of $e_O$ close to zero in the case $\theta_1=\pi/3$. Moreover,  larger values
of
$\theta_1$ allow larger ranges of $e_O$ with 
$m/|M|>1$ for any value of 
$\theta$.
For example, in the limiting case where supersymmetry is only broken
by the $F$--term of the five--brane $F^1$, i.e. $\theta_1=\pi/2$,
$m/|M|>1$ for $e_O>-0.65$. This case is shown in Fig.~\ref{susyonlyfivesoft},
where the soft terms and $m/|M|$ versus $e_O$ are plotted. Of course,
the figures are independent on $\theta$. Let us remark that 
scalar masses larger than gaugino masses are not easy to obtain,
as discussed below (\ref{gluinos1}), in the
weakly--coupled heterotic string.

It is worth noticing that the sum rule studied above, 
$3m^2=M^2$ for $\theta=\pi/6$ (and $\pi/6+\pi$) is no longer true 
when five--branes contributing to supersymmetry breaking are 
present.

%
\begin{figure}[htb]
\hspace{-1.0cm}
\epsfig{file= 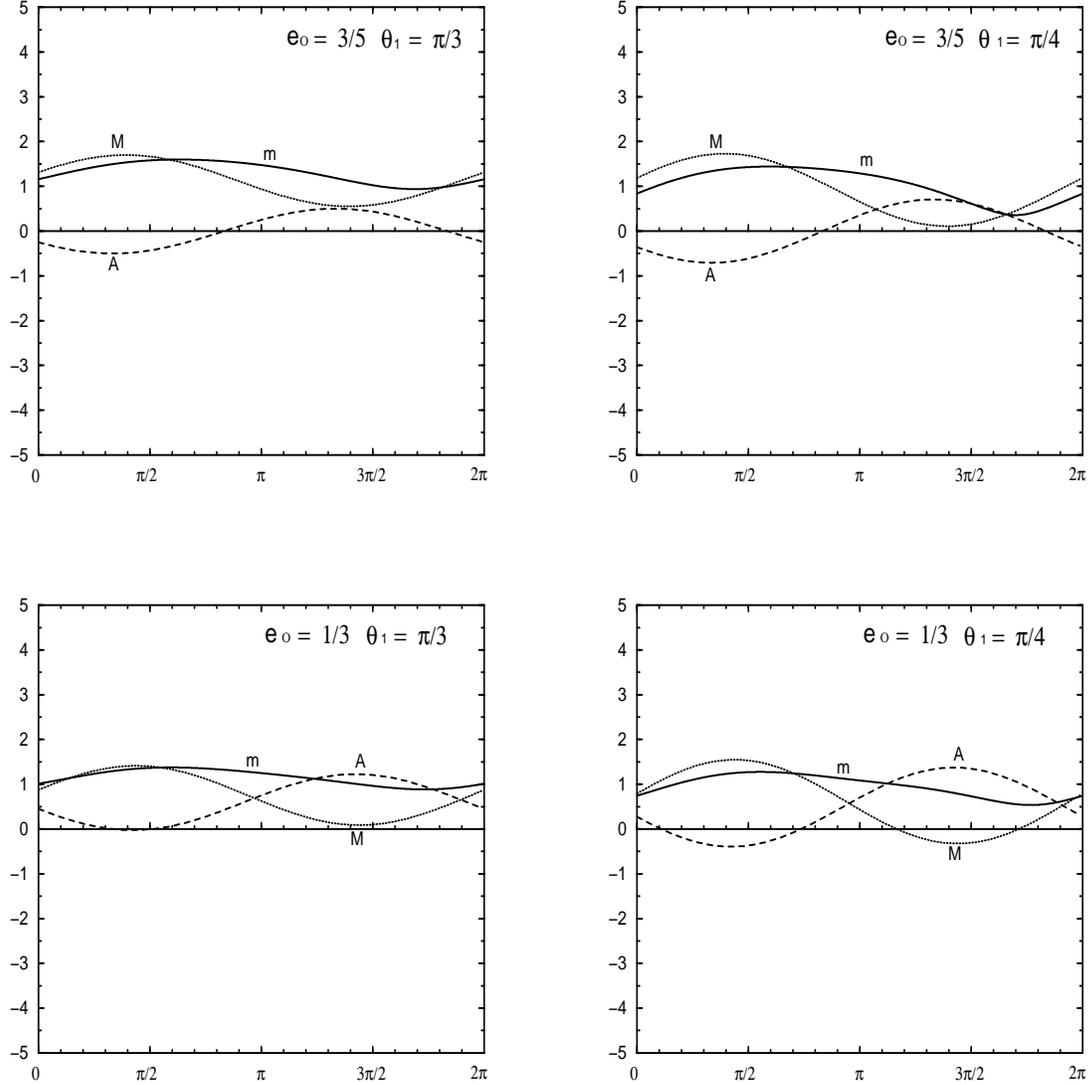, width=16cm, height=16cm}
\vspace{-1.0cm}
\caption{The same as Fig.~\ref{susyfivesoft} but for positive values 
of $e_O$.\label{susyfivesoft+}}
\end{figure}

Finally, let us mention that a different value for $\partial_1 K_5$
that
the one considered here, will only modify the expression of the $A$
parameter, since it is the only one which depends on that derivative.
On the other hand, one can check that 
a different value for $\partial_1\partial_{\bar 1} K_5$ 
will not modify the qualitative results concerning gaugino and scalar
masses.
Notice also that $\partial_1\partial_{\bar 1} K_5$ appears
always in the denominator in the expressions of the soft terms.
So, the larger the value of $\partial_1\partial_{\bar 1} K_5$ 
the smaller the deviation with respect to the case without five--branes
becomes.  
One can also check that  
different values for $z_1$, $\beta_O$ and $\beta_H$
will not modify the qualitative results concerning gaugino and scalar
masses.

%
\begin{figure}[htb]
\hspace{-0.7cm}
\epsfig{file= 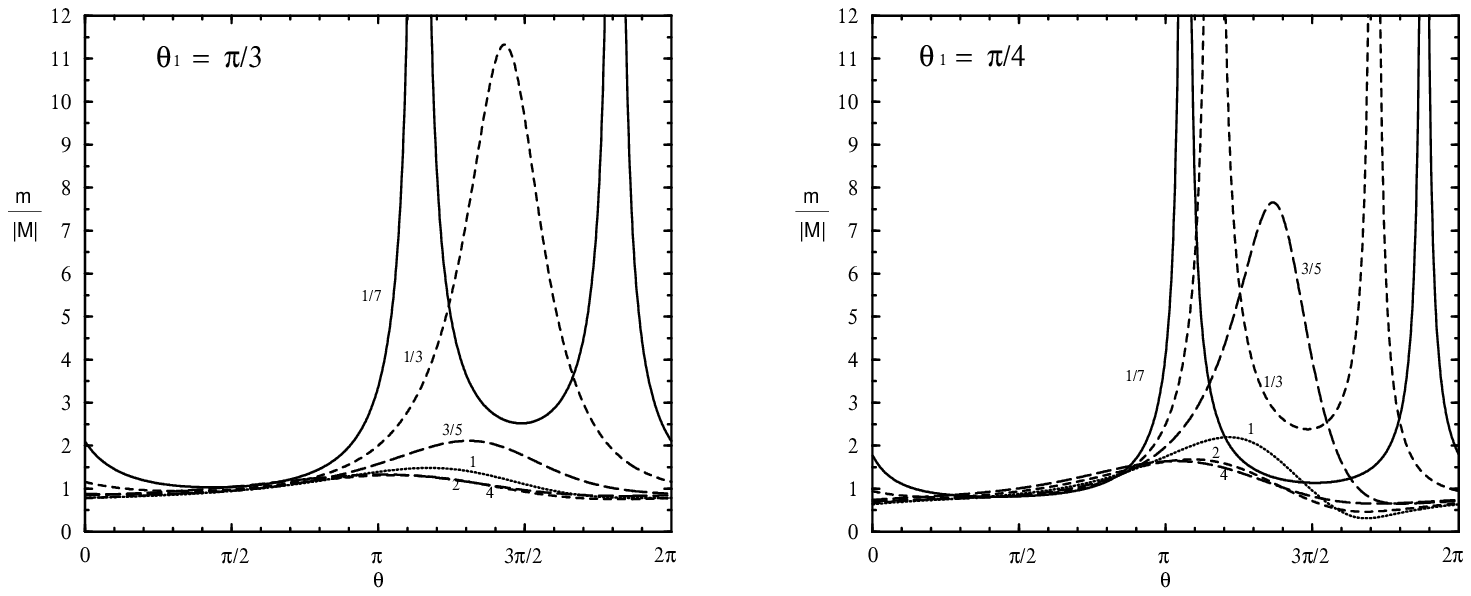, width=16cm, height=8cm}
\vspace{-1.0cm}
\caption{The same as Fig.~\ref{susyfivecociente} 
but for positive values of $e_O$.\label{susyfivecociente+}}
\end{figure}
%
%
\begin{figure}[htb]
\hspace{-0.7cm}
\epsfig{file= 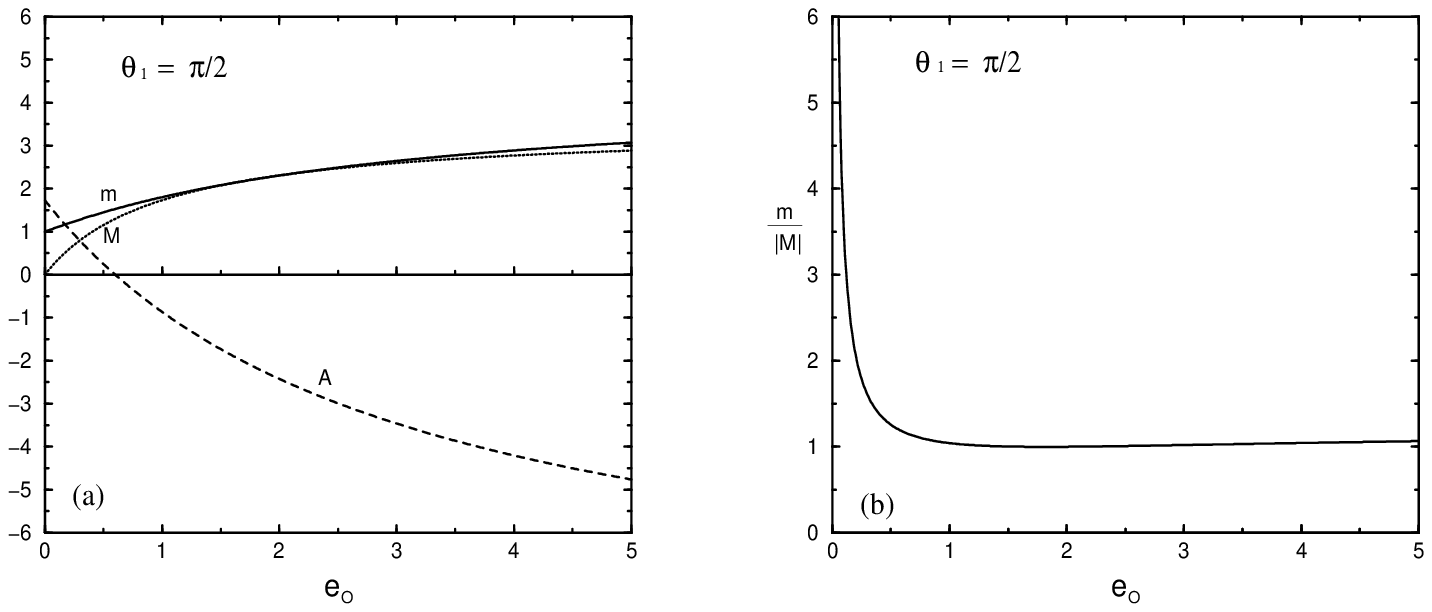, width=16cm, height=8cm}
\vspace{-1.0cm}
\caption{The same as Fig.~\ref{susyonlyfivesoft} 
but for positive values of $e_O$.\label{susyonlyfivesoft+}}
\end{figure}

We have not analyzed yet positive values of $e_O$ in the presence of
a five--brane contributing to supersymmetry breaking. In order to carry
it out we choose now $\beta_O=1$ and $\beta_1=-2$ which implies 
$b_O=b_H=1/2$. From (\ref{cota2}) in case {\it ii)} we deduce that the
whole range of allowed (positive) values of $e_O$ can be studied. We 
show in Fig.~\ref{susyfivesoft+} the soft terms for the values $e_O=1/3, 3/5$. 
For a fixed
value of $\theta_1$, the
larger
the value of $e_O$ the larger the range of soft terms becomes, unlike
the case without five--branes (see Fig.~1 in \cite{Mio}).
On the other hand, scalar masses larger than gaugino masses can also 
be easily obtained as in the case with negative values of $e_O$
studied above. This is plotted in more detail in
Fig.~\ref{susyfivecociente+} 
for different
values of $e_O$. For example, for $e_O=1/3$ and $\theta_1=\pi/3$,
$\theta\approx 3\pi/2$ one obtains $m/|M|\approx 10$.
Using (\ref{masas}) this result implies a relation of the type
(\ref{gluinos2}), 
$m_{\tilde l}\approx m_{\tilde q}\approx 3.5 M_{\tilde g}$.
In Fig.~\ref{susyonlyfivesoft+} we show the limiting case where
supersymmetry
is only broken by the $F$--term of the five--brane.

\section{Conclusions}

In the present paper we have tried to perform a systematic
analysis
of the soft supersymmetry breaking terms arising in a Calabi--Yau 
compactification
of the heterotic M--theory, as well as a detailed study of the
different scales of the theory. 
Since, as discussed in the introduction, 
Calabi--Yau manifolds with only
one K\"ahler modulus $T$ are very interesting from the phenomenological
point of view, not only because of their simplicity but also because
they might give rise to three--family models 
with an improvement with respect to the
non--universality problem, we have concentrated on these spaces. 

The soft
terms in the standard and non--standard embedding cases depend
explicitly on
the
gravitino mass $m_{3/2}$, the goldstino angle $\theta$ and the 
parameter $\epsilon_O$ (see (\ref{epsilonO})).
The only difference between both cases is the range of values where
$\epsilon_O$ is valid. $-1<\epsilon_O<1$ in the latter and $0<\epsilon_O<1$
in the former. This will give rise to different patterns of soft
terms.
In particular, scalar masses larger than gaugino masses in the
non--standard
embedding case are allowed (see Fig.~\ref{nonstandardcociente}), 
for narrow ranges of $\theta$, unlike
the standard embedding situation. 
This has obvious implications for low--energy ($\approx M_W$)
phenomenology, as discussed in (\ref{gluinos1}) and (\ref{gluinos2}).

The presence of non--perturbative objects as five--branes in the vacuum
modifies the previous analysis substantially. Even if the five--branes
do not contribute to supersymmetry breaking
the soft terms are modified. Basically, the soft terms are given
by the same formulae than in the standard and non--standard embedding 
but with a new parameter $e_O$ (instead of $\epsilon_O$), 
see (\ref{Es}), which is valid not only for $-1<e_O<1$ but also
for $e_O\geq 1$. However, although the pattern of soft terms is different,
scalar masses larger than gaugino masses are not possible 
(see Fig.~\ref{nonsusyfivecociente}) 
as in the
standard embedding situation. Other parameters appear when we allow
the
five--branes to contribute to supersymmetry breaking. In particular,
at least, a new goldstino angle $\theta_1$ must be included in the
computation of soft terms. In this way, scalar masses larger than
gaugino
masses can be obtained in a natural way (see
Figs.~\ref{susyfivecociente} 
and \ref{susyfivecociente+}). 
Depending on the values of 
$e_O$ and $\theta_1$, scalars could be heavier than gauginos 
even in the whole range of $\theta$. 
As discussed below (\ref{gluinos1}) this might be possible in the
weakly--coupled heterotic string only for $\sin\theta\rightarrow 0$.

Concerning the scales of the theory, we have discussed in detail the 
relations between the eleven--dimensional Planck mass, the
Calabi--Yau compactification
scale and the orbifold scale, taking into account higher order
corrections
to the formulae. Identifying 
the compactification scale with the GUT scale, it is easier
to obtain the phenomenologically favored value,
$M_{GUT} \approx 3\times 10^{16}$ GeV, in the
non--standard embedding than in the standard one 
(see Fig.~\ref{scales1}). On the other hand, to lower this scale 
(and therefore the eleven--dimensional Planck scale which is around two times
bigger)
to intermediate values $\approx 10^{11}$ GeV or $1$ TeV values or 
to obtain the radius of the orbifold as large as a millimetre 
is in principle
possible in some special limits. In particular, in the non--standard
embedding when $\epsilon_O\rightarrow -1$ and also in the case  
$\beta_O=0$ (see Fig.~\ref{scales2}).
However, the necessity of a fine--tuning in the former 
and the existence of a hierarchy problem in the latter 
render these possibilities
unnatural.
In the presence of five--branes, $M_{GUT}$ can be obtained more
easily (see Fig.~\ref{scales1five}a).
Although new possibilities arise in order to lower this scale, 
in particular when $e_O$ is very large (see Fig.~\ref{scales1five}b),
again at the cost of introducing a hughe hierarchy problem.

\bigskip

\noindent {\bf Note added}

\noindent As this manuscript was prepared, refs. \cite{Lii} and \cite{Kubo}
appeared. The former discusses some of the issues presented in this
work,
particularly the scenario of subsection 2.3.
The latter also discusses soft terms in the presence of five--branes,
however, its numerical analysis concentrates on the dilaton limit with
$0<e_O<2/3$.

\bigskip

\noindent {\bf Acknowledgments}

\noindent The work of D.G. Cerde\~no has been supported by a Universidad
Aut\'onoma de Madrid grant.
The work of C. Mu\~noz has been supported 
in part by the CICYT, under contract AEN97-1678-E, and
the European Union, under contract ERBFMRX CT96 0090.


\end{document}